\documentclass[11pt,a4paper]{article}
\pdfoutput=1
\usepackage{jheppub}
\usepackage{amsfonts}
\usepackage{amssymb}
\usepackage{amscd}

\title{A Unitary Model of The Black Hole Evaporation}

\author[a]{Yu-Lei Feng,}
\author[a,1]{Yi-Xin Chen,\note{Corresponding author.}}

\affiliation[a]{Zhejiang Institute of Modern Physics, Zhejiang University,\\
Hangzhou, 310027, P. R. China}
\emailAdd{yxchen@zimp.zju.edu.cn}

\abstract{
A unitary effective field model of the black hole evaporation is proposed to satisfy almost the four postulates of the black hole complementarity (BHC). In this model,
we enlarge a black hole-scalar field system by adding an extra radiation detector that couples with the scalar field. After performing a partial trace over the scalar field space, we obtain an effective entanglement between the
black hole and the detector (or radiation in it). As the whole system
evolves, the S-matrix formula can be constructed formally step by step. Without local quantum measurements, the paradoxes of the information loss and AMPS's firewall can be resolved.
However, the information can be lost due to quantum
decoherence, as long as some local measurement has been performed on the detector to acquire the information of the radiation in
it. But unlike Hawking's completely thermal spectrum, some residual correlations can be found in the radiations. All these considerations can be simplified in a qubit model that provides a \emph{modified quantum teleportation} to transfer the information via an EPR pairs.

}
\begin{document}
\maketitle
\flushbottom

\section{Introduction}
\label{sec:intro}

In 1976, Hawking pointed out the information loss paradox~\cite{a,b}, which says that the quantum state of the Hawking radiation emitted
from the black hole is not pure, but completely thermal. This means that the black
hole evaporation is not a unitary process, and information will be lost when the evaporation finishes. After Hawking's
work, various alternatives to restore the black hole unitary have
been proposed and studied, one of which is the black hole
complementarity (BHC)~\cite{c}. It postulates that

(i) \emph{black hole formation and evaporation are described via unitary
quantum evolution};

(ii) \emph{the region outside the stretched horizon is well described by
QFT in curved space};

(iii) \emph{to a distant observer, the black hole appears to be a quantum system with states given by, for example $\arrowvert
M\rangle$, with $M$ the mass of the black hole};

(iv) \emph{an in-falling observer can cross the horizon without
encountering any trouble, in particular, a field
vacuum can always be present in the near horizon region}.

However, in reference~\cite{d}, Almheiri et al formulated a
firewall paradox, saying that the postulates (i)(ii) and (iv) of the BHC are
not valid simultaneously. Their argument is briefly as follows: by dividing the Hawking radiation into an early part and
a late part, the purity of the Hawking radiation (postulate (i))
implies that the late part is fully entangled with the early
part, but the absence of any trouble for the in-falling
observer (postulate(iv)) implies that the late part is full entangled with the
modes behind the event horizon. This violates the monogamy of entanglement
of quantum mechanics. The firewall paradox has led to a serious
debate~\cite{d,e,f,g,h}.

Both of the two paradoxes seem to arise from a fact that their arguments depend too much on the observers. In the information loss argument, the distant observer plays an crucial role via local quantum measurements. While in the BHC, an in-falling observer is added to ensure the validness of effective field theory in the near horizon region, so that \emph{Einstein's equivalence principle} is satisfied. But this violates the monogamy of entanglement when combined with the descriptions of the distant observer. However, \emph{the principle of general covariance} says that a physical description should not depend on the observers. Thus, to reconcile the contradiction, the description of the distant observer should be extended to include a complementary interior observable\footnote{That is a combination $\{\mathcal{O}_{ext},\mathcal{O}_{int}\}$, for detailed discussions, see section~\ref{sec:bhc}.}. As a consequence, we can obtain an effective \emph{super-observer} whose description may be consistent with the in-falling observer's. In this case, a postulate about the interior region of the black hole is also needed, assuming that this region can also be well described by QFT in curved space, certainly the singularity $r=0$ should be excluded. This postulate seems to be appropriate only for a macro black hole whose interior region is large enough, but not for a micro one with a small interior, where quantum gravity effects will dominate. However, if this postulate can help to resolve those paradoxes for a macro black hole, it should be treated seriously.

The closed system in the black hole evaporation problem can be simplified to be
composed of two components, the black hole and a matter
field, for example, a scalar field. According to quantum mechanics~\cite{i}, a measurement apparatus is usually utilized as an
environment for a quantum measurement. It is thus possible to add an apparatus into the black hole evaporation problem, for example a radiation
detector that couples with the scalar field. With this added detector, an
effective field model can be proposed to satisfy the BHC, with the second postulate replaced by an extended one including the description about the interior of the black hole. After performing a partial trace over the scalar field space, an effective
black hole-detector (or radiation in it) entanglement will be obtained. The S-matrix formula can be
constructed formally step by step, implying that the information won't be lost during the
evolution of the entire system.

However, the information can be lost due to quantum decoherence, when some local measurement is
performed on the detector to acquire the information of the
radiation in it. In this sense, the black hole evaporation (without
extra matter absorptions) is analogous to the amplitude-damping
channel~\cite{i}, a schematic model of the decay of an excited atom
due to spontaneous emission of photons. The lowest order decay rate
is calculated, giving a smaller estimate of the $l=0$ luminosity (without
backscattering effects) $(128\pi^3 M^2)^{-1}$. Moreover, a
non-thermal spectrum differing from Hawking's completely thermal one
is also obtained, implying that the information is not completely
lost in this case.

A qualitative model including the gravitational perturbation is further investigated, in which the gravitons (or gravitational field
perturbations) play the role of an intermediate medium for the energy transfer between the interior and exterior of the black hole. Moreover, we show that the effective entanglements
between the black hole and the radiations (in the detector) belong to a class, whose members are nonlocal and generated by some other already existing entanglements.
This entanglement generation can be well demonstrated by a qubit model, in which correlations between two distant systems can be established through an EPR pairs\footnote{Readers who want to quickly understand our model conceptually are advised to read first the qubit model in section~\ref{sec:sum} and its mathematical details in appendix~\ref{sec:qubit}.}. Moreover, a \emph{modified quantum teleportation} via the EPR pairs is also proposed to transfer information, thus the information of the black hole can also be transferred outside effectively in our effective field model.

The outline of our paper is as follows. In section~\ref{sec:bhc}, we study the extension of the BHC (ii) to include a postulate about the interior region of a black hole. In sections~\ref{sec:model} and~\ref{sec:feature}, we develop our effective field model and obtain the required entanglement between the black hole and the added radiation detector. In section~\ref{sec:S}, a S-matrix formula for our model is constructed, while in sections~\ref{sec:infor} and~\ref{sec:mix}, the situations of the information for our model is discussed in detail. The inclusion of the gravitational perturbation is qualitatively analyzed in section~\ref{sec:gra}. Finally, in section~\ref{sec:sum}, we give a brief summary and propose a qubit model of the black hole evaporation. Two appendixes are added. In appendix~\ref{sec:singular}, a simple model with a singular evolution operator is studied, and in appendix~\ref{sec:qubit}, the mathematical detail of the qubit model is given.

\section{Effective Field Model of The Black Hole Evaporation}
\label{sec:Model}

\subsection{General Covariance and Extension of the BHC (ii)}
\label{sec:bhc}

In the physics of a black hole, for example, a Schwarzschild black hole with a mass $M$, there are mainly two classes of observers: one class consists of the distant observers, or more generally static observers, while the other one is composed of the in-falling observers. For a static observer, the reference frame is given by the global (Schwarzschild) coordinate $(t,r,\theta,\phi)$ with a metric singularity at $r=2M$, leading to a hypersurface called an event horizon. As a result, in the view of a distant observer, an in-falling particle will never cross the event horizon to enter the interior of the black hole, in the sense of using infinite time because of the singular event horizon at $r=2M$; analogously, a beam of light in the interior of the black hole can never escape outside by crossing the event horizon because of the same singular event horizon. While for an in-falling observer, the chosen coordinate is some locally inertial one $\xi^{\alpha}_{X}$ so that no singularity occurs. Consequently, in his view, the in-falling particle can cross the event horizon in a finite time.

The descriptions of the above two classes of observers are apparently in contradiction, which seems to invalidate \emph{the principle of general covariance}. In the quantum version, this contradiction is in fact expressed as the information loss paradox, in which the quantum state is mixed in the view of a distant observer, while pure for an in-falling observer. Certainly, the BHC was actually proposed to reconcile the contradiction between the two descriptions, but it seems to violate the monogamy of entanglement
of quantum mechanics, leading to a firewall paradox~\cite{d}. Notice further that the firewall paradox arises because of the possible inconsistency of the BHC(i)(ii)(iv), with (ii)(iv) involving the static and in-falling observers respectively. In this sense, the firewall paradox can also be treated as one part of the violation of general covariance. In reference~\cite{h}, by adding an ancillary Hilbert space, the authors tried to reconstruct the local effective field theory observables that probe the black hole interior, and relative to which the state near the horizon looks like a local Minkowski vacuum. In this way, the firewall paradox can be resolved effectively, but not completely. Moreover, they treated their ancillary Hilbert space only as a carbon copy of the space of the exterior Hawking radiation. Then whether their ancillary Hilbert space can be treated just as the space of the interior modes?

Let's skip this problem temporarily, and compare the descriptions of two kinds of observers in a different way. If the principle of general covariance is also proper in the quantum version, the descriptions of two different observers about the physical world should be consistent with each other. However, this seems not to be the case for the distant and in-falling observers in a black hole background or the static and accelerated observers in a flat space background, indicated by the information loss paradox. Let's consider a simplified model of formation of a black hole, a shock wave model~\cite{j}. Initially, the space-time is a flat one, and after a while, a black hole is produced by a shock wave. The resulting black hole can be well described by the Vaidya space-time with line element~\cite{j}
\begin{equation}
\label{eq:a}
\begin{split}
ds^2=-(1-\frac{M(v)}{r})dv^2+2dvdr+r^2d\Omega^2\,,
\end{split}
\end{equation}
with $M(v)=M\Theta(v-v_0)$. For a static observer relative to the initial flat space-time, an initial pure state will evolve to a mixed state due to the Hawking effect~\cite{a}, i.e. the familiar information loss paradox. For an accelerated observer relative to the initial flat space-time, however, the initial state he observe should be thermal due to the Unruh effect~\cite{j1}, since he is accelerated relative to the initial flat space-time. Moreover, if his acceleration can be treated as the one induced by the produced black hole according to Einstein's equivalence principle, i.e. he becomes an in-falling observer for the formed black hole, then the final state he observes may be pure\footnote{More concretely, the coordinate frame of a static observer is $(t,r,\theta,\phi)$, which is \emph{inertial} for the initial flat space-time, but \emph{non-inertial} for the final produced black hole. However, the coordinate frame of an accelerated observer is some one $\xi^{\alpha}_{X}$, which is \emph{non-inertial} for the initial flat space-time. But it may be \emph{inertial} for the final black hole if $\xi^{\alpha}_{X}$ is just the in-falling coordinate frame of the formed black hole, according to Einstein's equivalence principle.}. These can be expressed formally as
\begin{equation}
\label{eq:b}
\begin{split}
\arrowvert
0\rangle_{in}\stackrel{static}{\longrightarrow}\rho_{mix},\qquad \rho_{mix}\stackrel{in-falling}{\longrightarrow}\arrowvert
0\rangle_{U}\,,
\end{split}
\end{equation}
where $\arrowvert0\rangle_{in}$ is the Minkowski vacuum for the static observer relative to the initial flat space-time, and $\arrowvert
0\rangle_{U}\sim\arrowvert0\rangle_{in}$ is the Unruh (or near horizon) vacuum for the in-falling observer relative to the produced black hole. Then how to understand these processes, especially the second one involving \emph{an evolution from a mixed state to a pure one}? If the first process was physically possible, then the second one should also be the case. Although the first process can be obtained by some partial trace in quantum mechanics, the second one can not be realized effectively. These two inconsistent processes observed by two different observers in fact imply further the violation of the principle of general covariance, in addition to the information loss paradox\footnote{Since the information loss can be explained well by a local measurement according to quantum mechanics, the relevance to the violation of general covariance is not evident. While the second process in~\eqref{eq:b} provides an obvious evidence for the violation of general covariance, since it cannot be explained well only by a local measurement.}.

The processes in~\eqref{eq:b} are both non-unitary, because some local measurements have been performed either initially or finally, indicated by the mixed density operators. Obviously, these local measurements are caused by the space-time causal structures in the views of respective observers. It thus seems that the only way to reconcile the contradiction is to extend those local measurements(or observables) by including some complementary ones. This is analogous to the arguments of reference~\cite{h}, resolving the firewall paradox effectively by adding an ancillary Hilbert space. \emph{By treating this ancillary Hilbert space just as the one for the interior modes}, the descriptions of the static and in-falling observers may be consistent with each other, and the principle of general covariance may thus be obeyed\footnote{Notice that the violation of general covariance is expressed by both of the paradoxes of information loss and firewall. The authors of reference~\cite{h} only deal with the firewall paradox, since they still base their arguments on a thermal spectrum, i.e. the Hawking's process in~\eqref{eq:b}. In this sense, the violation of general covariance is resolved only partially by including a complementary observable, with the information loss paradox still unresolved. This can also be seen from the discussions below~\eqref{eq:f}.}. In other words, according to the principle of general covariance, there should be a correspondence between the field observables relative respectively to the in-falling and static observers expressed formally as
\begin{equation}
\label{eq:c}
\begin{split}
\{\mathcal{O}_{in-falling}\}\rightleftharpoons \{\mathcal{O}_{ext},\mathcal{O}_{int}\}\,,
\end{split}
\end{equation}
with $O_{ext}$ and $O_{int}$ the observables corresponding to the static observers in the exterior and interior of the black hole respectively\footnote{In the flat space background, there is an analogous correspondence $\{\mathcal{O}_{static}\}\rightleftharpoons \{\mathcal{O}_{R},\mathcal{O}_{L}\}$ between the observables relative respectively to the static and accelerated observers.}. As will be shown below, this correspondence implies a possible global or nonlocal correlation between $O_{ext}$ and $O_{int}$, meaning that a single $O_{ext}$ or $O_{int}$ is not complete enough to give a full consistent description.

Then whether the violation of general covariance can be resolved by the correspondence~\eqref{eq:c}? Actually, this correspondence indicates that $\{\mathcal{O}_{ext},\mathcal{O}_{int}\}$ may be treated as the ``super-observer'' of~\cite{c}, whose description can involve both the interior and exterior degrees of freedom. This resembles the case of an EPR pairs, with the identifications
\begin{equation}
\label{eq:e}
\begin{split}
\mathcal{O}_{in-falling}\Rightarrow S_t=S_1+S_2,\qquad \mathcal{O}_{ext}\Rightarrow S_1,\mathcal{O}_{int}\Rightarrow S_2 \,,
\end{split}
\end{equation}
where $S_t$ is the total spin of two electrons with respective spins $S_1$ and $S_2$. In other words, the observables $O_{ext}$ and $O_{int}$ are independent and can be measured simultaneously, but the measurement results may be correlated for a pure state of the in-falling observer, corresponding to case of the total spin states or the four Bell states. This means that, for the Unruh vacuum $\arrowvert0_{U}\rangle$ we have
\begin{equation}
\label{eq:f}
\begin{split}
\langle 0_{U}\arrowvert\mathcal{O}_{ext}\mathcal{O}_{int}\arrowvert0_{U}\rangle\neq \langle 0_{U}\arrowvert \mathcal{O}_{ext}\arrowvert0_{U}\rangle\langle 0_{U}\arrowvert \mathcal{O}_{int}\arrowvert0_{U}\rangle\,.
\end{split}
\end{equation}
For the case of EPR pairs, the correlations implicit in a Bell state can be acquired by first locally measuring $S_1$ and $S_2$ then comparing the measurement outcomes via information transfer through a classical channel\footnote{A quantum channel seems to be more efficient~\cite{i}, and it can easily be established in a flat space-time. In fact, our effective field model indeed establishes a quantum channel between the causally disconnected interior and exterior of the black hole, by using of the entanglement implicit in the (near horizon) vacuum state $\arrowvert0_{U}\rangle$. This can also be quickly seen from the qubit model proposed in section~\ref{sec:sum} and appendix~\ref{sec:qubit}. As a result, the correlations of~\eqref{eq:f} may be acquired by comparing the measurement outcomes through a quantum channel established via another field's near horizon vacuum state. However, not all of the field's correlations can be acquired in this way, since local measurements always destroy correlations. Thus the only resolution is a complete quantum gravity theory without background dependence, as argued below. }, obtaining the correlation function $\langle S_1S_2\rangle$. However in acquiring the correlations of~\eqref{eq:f}, local measurements can be performed in principle while the comparison cannot be achieved because of the causal disconnectedness of the exterior and interior of a black hole. This means that the ``super-observer'' $\{\mathcal{O}_{ext},\mathcal{O}_{int}\}$ can not be realized physically, which may be treated as another (stronger) version of information loss. Since the information loss paradox cannot be resolved in this way, it may be concluded that, \emph{general covariance is violated in the framework of effective field theory with dependent space-time background}. The only resolution may be the unknown complete quantum gravity theory without background dependence.

Although the correlation information of~\eqref{eq:f} cannot be acquired \emph{through local measurements} in the background of a black hole, that (nonlocal) correlation is indeed present, implicit in the near horizon vacuum state. Since the (correlation) information loss problem is mainly caused by the causal structure of the black hole, what we have to do is to \emph{make the black hole evaporate completely in a (effective) unitary manner according to effective field theory, without involving any local measurement}. After the black hole disappear, there will not be mode splitting, and the problem of (correlation) information loss is absent. This means that we should develop an effective unitary method to transfer the energy of the black hole outside \footnote{This is analogous for the accelerated observer in a flat space-time, whose acceleration is provided by some extra (energy) source that resembles to a black hole when expressed as a curved metric. The reason for `` effective" is due to the (classical) background dependence of the effective field theory, i.e. a prior local measurement $\langle \hat{g}_{\mu\nu}\rangle$. A full unitary description can be obtained only through the quantum gravity theory. }. In Hawking's argument, the semiclassical Einstein equation for only the exterior region is used, where local measurement $\langle 0_{U}\arrowvert\mathcal{O}_{ext}\arrowvert0_{U}\rangle$ serves as the source term. Therefore, by means of this semiclassical method, the (correlation) information will always be lost. We shall show in this paper that, a unitary effective field model can be proposed to transfer energy or information between the interior and the exterior of the black hole, by using of the nonlocal correlation~\eqref{eq:f}, or more precisely the entanglement implicit in the near horizon vacuum state. To apply that nonlocal correlation, or to resolve the firewall paradox implicit in the BHC, the BHC (ii) should be extended as:

(ii') \emph{both of the exterior and interior regions of the black hole can be well described by
QFT in curved space, with the singularity $r=0$ excluded from the interior region}.

That is to say, there are two effective field theories on both the two sides of the event horizon, which are independent from each other in the sense that the observables are constructed with different modes of the fields. This is such a crucial extension that an effective field model can be proposed to satisfy the \emph{extended} BHC(i)(ii')(iii)(iv), so that the black hole can evaporate completely in a (effective) unitary manner.

\subsection{The Effective Field Model}
\label{sec:model}

\subsubsection{The Hilbert Space: Introduction of a Radiation Detector}
\label{sec:hilbert}

As analyzed in the last subsection, in the black hole background locally measuring $\mathcal{O}_{ext}$ (or $\mathcal{O}_{int}$) will always lead to information loss due to the black hole's causal structure. Then how to avoid these local measurements? According to quantum mechanics~\cite{i}, a quantum measurement should be performed via a coupling between the target system and a measurement apparatus. In the black hole evaporation problem, we can also add a radiation detector that couples with the target system, for example, a scalar field. This detector can be located somewhere as a static observer, or can fall freely into the black hole as an in-falling observer. In our model, we treat it as a distant static observer.
This radiation detector works in the following way. By coupling with the scalar field in the exterior of the black hole, the detector will be entangled with the exterior modes of the scalar field, so that a local measurement or observable $\mathcal{O}_{ext}$ can be replaced by another one $\mathcal{O}_{D}$ performed on the space of the detector. This approach is different from the standard one utilized in quantum measurement theory\footnote{Our coupling between the scalar field and the detector is constructed according to effective field theory, as given in~\eqref{eq:13}. The differences between the two couplings will be shown in section~\ref{sec:feature}.}. In fact, what we need is just \emph{an environment that interacts with the scalar field}, so that the energy or information can be transferred between them. Unlike the space of the scalar field which has been split into two parts due to the black hole's causal structure, the space of the detector is complete enough since the detector is located completely in the exterior of the black hole in our model.

Assume the extended BHC(with postulate (ii') given in the last subsection) is proper. The
total Hilbert space of the entire system, including a (Schwarzschild) black hole
$B$, a scalar field $\psi$ and an added perfect radiation detector
$D$ (without energy loss), can be factorized as
\begin{equation}
\label{eq:1}
\begin{split}
\mathcal {H} &= \mathcal {H}_B\otimes\mathcal
{H}_{\psi}\otimes\mathcal {H}_D \,,
\end{split}
\end{equation}
where the space of the scalar field $\mathcal
{H}_{\psi}$ is composed of exterior and interior modes. From the BHC (iii) and (iv), however, we can restrict our considerations
within a smaller space
\begin{equation}
\label{eq:2}
\begin{split}
\mathcal {H}_0 &= \mathcal {H}_{B,\psi}\otimes\mathcal {H}_D \,,
\end{split}
\end{equation}
in which a state can be expressed as\footnote{Here $M_0\simeq M+E$ is the total energy of the closed system.}
\begin{equation}
\label{eq:3}
\begin{split}
\arrowvert\Phi\rangle
&=(\sum_{M,E\geq0}^{M_0}C(M)\arrowvert
M\rangle_{B}\arrowvert0_{M}\rangle_{\psi})\otimes\arrowvert
E\rangle_D\,,
\end{split}
\end{equation}
where $\arrowvert M\rangle_{B}$ and $\arrowvert E\rangle_D$ are the
orthonormal basis states of the black hole and the detector,
respectively. The entangled state in the bracket describes partially
the correlation\footnote{The full correlation is actually described
by the coupling between the black hole and the scalar field. Here, the
entangled state in the bracket just gives the space-time
background dependence of the scalar field, with the back-reaction of the scalar field on the black hole being
modeled in~\eqref{eq:7} below.} between the black hole and the scalar field. $\arrowvert0_{M}\rangle_{\psi}$ or $\arrowvert0_{M}\rangle$ for short (with $M$ denoted as the
dependence on black hole's mass\footnote{This mass dependence
makes the problem a little complicated. In this section, we ignore
its possible effects, which will be studied in section~\ref{sec:dynamics}.}) is the near horizon vacuum of the scalar field~\cite{j}
\begin{equation}
\label{eq:4}
\begin{split}
\arrowvert0_{M}\rangle_{\psi} &= \prod_\omega(1-e^{-8\pi
M\omega})^{1/2}\exp(\sum_{\omega}e^{-4\pi
M\omega}b_{\omega}^{\dag}\tilde{b}_{\omega}^{\dag})\arrowvert0,\tilde{0}\rangle\,,
\end{split}
\end{equation}
with $b_{\omega}^{\dag}$ and $\tilde{b}_{\omega}^{\dag}$ the creators of exterior and interior modes respectively. Here, the formula for the initial vacuum state in the shock wave model is used, since $\arrowvert0\rangle_{in}\sim \arrowvert0\rangle_{U}$~\cite{j}. The state $\arrowvert0\rangle_{B}$ in~\eqref{eq:3} only stands for a space-time without black holes, with $\arrowvert0_0\rangle_{\psi}$ the corresponding scalar field vacuum. The free
Hamiltonian of the entire system is
\begin{equation}
\label{eq:5}
\begin{split}
H_0 &= H_B+H_{\psi}+H_D \,,
\end{split}
\end{equation}
with $H_D$ chosen simply as
\begin{equation}
\label{eq:6}
\begin{split}
H_D &=
\sum_{\tilde{\omega}}\tilde{\omega}d_{\tilde{\omega}}^{\dag}d_{\tilde{\omega}}
\,,
\end{split}
\end{equation}
where $d^{\dag}_{\tilde{\omega}}$ and $d_{\tilde{\omega}}$ stand for the
raising and lowering operators of the energy levels of the
detector. As for the $H_{\psi}$, it can be derived from a general formula
\begin{equation}
\label{eq:6a}
\begin{split}
H_{\psi}(t) &= \int_{\Sigma_t}T_{\alpha\beta}K^{\alpha}d\Sigma^{\beta}
\,,
\end{split}
\end{equation}
with a time translation Killing field $K^{\alpha}$ on a space-like Cauchy hypersurface $\Sigma_t$. When $\Sigma_t$ approaches the infinite past $I^-$, i.e. $\Sigma_t\rightarrow I^-$, the corresponding free Hamiltonian becomes
\begin{equation}
\label{eq:6b}
\begin{split}
H_{\psi}(t) \stackrel{I^-}{\longrightarrow} H_{a}=\sum_{\omega}\omega a_{\omega}^{\dag}a_{\omega}
\,,
\end{split}
\end{equation}
where $a_{\omega}\arrowvert0_{M}\rangle_{\psi}=0$. For a general $\Sigma_t$ intersecting the event horizon, it will be split into $\Sigma_{ext}\bigcup\Sigma_{int}$. When it approaches $I^+\bigcup H^+$, i.e. the infinite future together with the future event horizon, the free Hamiltonian will be given by
\begin{equation}
\label{eq:6c}
\begin{split}
H_{\psi}(t) &= \int_{\Sigma_{ext}}T_{\alpha\beta}K^{\alpha}d\Sigma^{\beta}+\int_{\Sigma_{int}}T_{\alpha\beta}K^{\alpha}d\Sigma^{\beta}\stackrel{I^+\bigcup H^+}{\longrightarrow} H_{b}+H_{\tilde{b}}
\,.
\end{split}
\end{equation}
Notice that the state $\arrowvert
M\rangle_{B}\arrowvert0_{M}\rangle_{\psi}\otimes\arrowvert
E\rangle_D$ in~\eqref{eq:3} is an eigenstate of the free Hamiltonian in~\eqref{eq:5}, with $H_{\psi}$ given by $H_{a}$ in~\eqref{eq:6b}, this property can be utilized to construct the S-matrix, as shown in section~\ref{sec:S}.

\subsubsection{The Interior and Exterior Interactions}
\label{sec:interaction}

To verify the consistency of the extended BHC, we have to evolve the entire
system and see whether it can always be descried well within the
smaller space $\mathcal {H}_0$. An interaction term $H_{int}(t)$
is needed. The details of the interaction may involve some unknown quantum gravity effects, but we
can still propose a simple model based on effective field theory. The scalar field can diffuse over the whole space-time, including both the exterior and interior of the black hole. Then, for a static observer, according to the extended BHC (ii') given in section~\ref{sec:bhc}, the full interaction can be chosen as $H_{int}(t)=H_{B,\psi}(t)+H_{\psi,D}(t)$. The term $H_{B,\psi}(t)$ gives the interaction between the black hole and the scalar field, which is localized in the interior of the event horizon, while $H_{\psi,D}(t)$ is a local interaction between the scalar field and the detector in the exterior of the event horizon. Moreover, these two terms should be independent from each other because of the causal structure of the black hole. However, we shall show below that the entanglements, between the $b_{\omega}$ and $\tilde{b}_{\omega}$ modes implicit in the vacuum state $\arrowvert0_{M}\rangle$, can be applied to correlate the causally disconnected interior and exterior of the black hole, which leads to the evaporation of the black hole.

Under these circumstances, $H_{B,\psi}(t)$ could simply be chosen as a direct coupling between the scalar field and the black hole
\begin{equation}
\label{eq:7}
\begin{split}
H_{B,\psi}(t) &= \int_{\Sigma_{int}^t} d^3x\psi(t)\mathbf{V}(t) \,,
\end{split}
\end{equation}
where $\mathbf{V}(t)$ is a space-distribution operator acting on the Hilbert space of the black hole, and the interaction region $t\times \Sigma_{int}^t$ is in the interior of
the black hole(or event horizon). This interaction term is different from the one used in~\cite{h}, where the interaction happens only at the stretched horizon, or in the exterior of the event horizon. Here according to the extended BHC (ii'), the interaction between the black hole and the scalar field are assigned to happen in the interior of the event horizon. We can still expand
the operator $\mathbf{V}(t)$ in terms of
$\mathbf{V}_{\omega}^{\dag}$ and $\mathbf{V}_{\omega}$ that map
black hole states $\arrowvert M\rangle_{B}$ to $\arrowvert
M\pm\omega\rangle_{B}$, together with the field vacua
$\arrowvert0_M\rangle_{\psi}$ to
$\arrowvert0_{M\pm\omega}\rangle_{\psi}$ due to the correlation
in~\eqref{eq:3}. If concerning with only the black hole evaporation,
$V_{M,\omega}\equiv V_{M-\omega,M}(\omega)=\langle
M-\omega\arrowvert\mathbf{V}_{\omega}\arrowvert M\rangle$ is the
required matrix element, determining the emission rate of particle
of frequency $\omega$. To obtain these matrix elements, one can use an approximate completeness
relation for the restricted space $\mathcal {H}_{B,\psi}$\footnote{For a state $\arrowvert\Phi\rangle=\sum_{i}c_i\arrowvert
i\rangle\arrowvert\phi_i\rangle$, with an orthonormal basis
$\langle j\arrowvert i\rangle=\delta_{ij}$ and normalized but
non-orthogonal states $\langle\phi_i\arrowvert\phi_i\rangle=1$, we
have $\sum_i\arrowvert i\rangle\langle
i|\otimes\arrowvert\phi_i\rangle\langle\phi_i\arrowvert\arrowvert\Phi\rangle=\arrowvert\Phi\rangle$,
meaning that $\sum_i\arrowvert i\rangle\langle
i\arrowvert\otimes\arrowvert\phi_i\rangle\langle\phi_i\arrowvert$ is
an identity-like operator. However, in a general space this relation
is not fulfilled.}
\begin{equation}
\label{eq:8}
\begin{split}
\mathbb{I} &= \sum_{M}\arrowvert M\rangle_B\langle
M\arrowvert\otimes\arrowvert 0_M\rangle_{\psi}\langle 0_M\arrowvert \,.
\end{split}
\end{equation}
In terms of the modes in the stationary regions, the scalar field can be expanded as
\begin{equation}
\label{eq:9}
\begin{split}
\psi &=  \int_{0}^{\infty}
d\omega(a_{\omega}^{\dag}U_{\omega}^*+h.c.) = \int_{0}^{\infty}
d\omega(b_{\omega}^{\dag}u_{\omega}^*+\tilde{b}_{\omega}\tilde{u}_{\omega}+h.c.)
\,,
\end{split}
\end{equation}
with $U_{\omega}\sim (\omega^{1/2}r)^{-1}e^{-i\omega v}$ the ingoing modes at the infinite past $I^-$. The outgoing modes
$u_{\omega}$ at the infinite future $I^+$ and the incoming modes $\tilde{u}_{\omega}$ at the
future horizon $H^+$ are~\cite{j}
\begin{subequations}\label{eq:10}
\begin{align}
\label{eq:10:1} u_{\omega}=
-\frac{1}{4\pi\sqrt{\omega}}\frac{\exp[-i\omega(v_H-4M\ln\frac{v_H-v}{4M})]}{r}\theta(v_H-v)
\,,
\\
\label{eq:10:2}
\tilde{u}_{\omega}=-\frac{1}{4\pi\sqrt{\omega}}\frac{\exp[i\omega(v_H-4M\ln\frac{v-v_H}{4M})]}{r}\theta(v-v_H)
\,.
\end{align}
\end{subequations}
Here we still use the formulae for the shock wave model, and consider only
the s-wave components without the backscattering effects for simplicity. By substituting~\eqref{eq:10} into~\eqref{eq:7}, and noting that the
interaction region is in the interior of the black hole, we have
\begin{equation}
\label{eq:11}
\begin{split}
H_{B,\psi}(t) &= \sum_{\omega}[\mathbf{V}_{\omega}(t)\tilde{b}_{\omega}+h.c.] \,,
\end{split}
\end{equation}
with $\mathbf{V}_{\omega}(t)\sim \int
d^3x\mathbf{V}(t)\tilde{u}_{\omega}$. Then, by using of~\eqref{eq:8}, we have
\begin{equation}
\label{eq:12}
\begin{split}
\sum_{M\geq\omega,\omega}V_{M,\omega}\arrowvert
M-\omega\rangle_B\langle
M|\otimes\arrowvert0_{M-\omega}\rangle_{\psi}\langle0_{M-\omega}\arrowvert\tilde{b}_{\omega}\arrowvert0_M\rangle_{\psi}\langle0_M\arrowvert
\,,
\end{split}
\end{equation}
where the factor
$_{\psi}\langle0_{M-\omega}\arrowvert\tilde{b}_{\omega}\arrowvert0_M\rangle_{\psi}$
vanishes according to~\eqref{eq:4}, i.e. out of the space $\mathcal {H}_0$.

To avoid this, the created particle needs to be transported somewhere else, for example into the radiation detector. This can be accomplished by the
interaction $H_{\psi,D}(t)$
\begin{equation}
\label{eq:13}
\begin{split}
H_{\psi,D}(t) &= \tilde{g}\int_{\Sigma_{ext}^t} d^3x\psi(t)\phi_D(t)= \sum_{\omega}[g_{\omega}(t)b_{\omega}
d_{\omega}^{\dag}+h.c.]\,,
\end{split}
\end{equation}
where $g_{\omega}(t)d_{\omega}^{\dag}\sim \tilde{g}\int d^3x\phi_D(t)u_{\omega}$. $\phi_D$ stands for some (localized) field inside the detector, and the
interaction region $t\times \Sigma_{ext}^t$ is near the infinite future
$I^+$. The unitary evolution operator for the full interaction is
\begin{equation}
\label{eq:14}
\begin{split}
U_{B,\psi,D} &= e^{-i\int dtH_{int}(t)}=\exp i\Big\{\sum_{\omega}\int
dt[g_{\omega}(t)b_{\omega}
d_{\omega}^{\dag}+\mathbf{V}_{\omega}(t)\tilde{b}_{\omega}]+h.c.\Big\}
\,,
\end{split}
\end{equation}
where the emission and absorption parts have been grouped separately, while the relevant terms for the evaporation process is
\begin{equation}
\label{eq:15}
\begin{split}
\exp i\sum_{\omega}\int dt[g_{\omega}(t)b_{\omega}
d_{\omega}^{\dag}+\mathbf{V}_{\omega}(t)\tilde{b}_{\omega}] \,.
\end{split}
\end{equation}
Then instead of the vanishing factor in~\eqref{eq:12}, we will
obtain a non-vanishing one
\begin{equation}
\label{eq:16}
\begin{split}
N(M,\omega)\equiv \langle0_{M-\omega}\arrowvert
b_{\omega}\tilde{b}_{\omega}\arrowvert0_M\rangle=e^{-4\pi
M\omega}\langle0_{M-\omega}\arrowvert
b_{\omega}b_{\omega}^{\dag}\arrowvert0_M\rangle  \,,
\end{split}
\end{equation}
where the relation $\tilde{b}_{\omega}\arrowvert0_M\rangle=e^{-4\pi
M\omega}b_{\omega}^{\dag}\arrowvert0_M\rangle$~\cite{j} has been
used. Considering the operation of~\eqref{eq:15} on
$\arrowvert\Phi\rangle$ in~\eqref{eq:3}, the lowest order term is given as
\begin{equation}
\label{eq:17}
\begin{split}
\sum_{M\geq\omega;\omega,E\geq 0}^{M_0}-\{N(M,\omega)g_{\omega}V_{M,\omega}\}C(M)
\arrowvert
M-\omega\rangle_{B}\arrowvert0_{M-\omega}\rangle_{\psi}\otimes
d_{\omega}^{\dag}\arrowvert E\rangle_D \,,
\end{split}
\end{equation}
which expresses an entanglement or correlation between the black hole and
the radiation in the detector. This entanglement between the causally disconnected interior and exterior of the black hole, is generated by the entanglements between the $b_{\omega}$ and $\tilde{b}_{\omega}$ modes implicit in the vacuum $\arrowvert0_{M}\rangle_{\psi}$. A more detailed discussion about this entanglement will be given in section~\ref{sec:infor}. If the black hole continues evaporating, higher order terms will contribute. Without extra
matter absorptions, for an initial state $\arrowvert
M_0\rangle_{B}\arrowvert0_{M_0}\rangle_{\psi}\otimes\arrowvert0\rangle_D$,
the black hole may evaporate completely in the end, leading to a
final state\footnote{~\eqref{eq:18} is not the only final state, since the
interaction $H_{\psi,D}(t)$ can correlate the scalar field with the detector. If the absorption part in
\eqref{eq:14} is included, a state $\arrowvert
M_{min}\rangle_{B}\arrowvert0_{M_{min}}\rangle_{\psi}\otimes\arrowvert
E_{max}\rangle_D$ may also be obtained, i.e. a dynamic balance with a ``remnant'' in the black hole.}
\begin{equation}
\label{eq:18}
\begin{split}
\arrowvert0\rangle_{B}\arrowvert0_{0}\rangle_{\psi}\otimes\arrowvert
M_0\rangle_{D}(t\rightarrow+\infty) \,,
\end{split}
\end{equation}
which is still in the restricted space $\mathcal {H}_0$. These two states can
thus be related to each other by an S-matrix, which will be shown in
section~\ref{sec:S}.

Now let's consider the factor $N(M,\omega)$ defined in~\eqref{eq:16}. From~\eqref{eq:17} this factor is a part of the emission amplitude for the evaporation. It is formally similar to the expectation value of the particle number
$\bar{N}(M,\omega)=\langle0_{M}\arrowvert
b_{\omega}^{\dag}b_{\omega}\arrowvert0_M\rangle$. To compare them, let's
calculate another quantity
$\tilde{N}(M,\omega)=\langle0_{M-\omega}\arrowvert
b_{\omega}^{\dag}b_{\omega}\arrowvert0_M\rangle$. After some
calculations we have
\begin{equation}
\label{eq:19}
\begin{split}
\tilde{N}(M,\omega) = \prod_{\omega'\neq\omega}\frac{(1-e^{-8\pi
M\omega'})^{1/2}(1-e^{-8\pi (M-\omega)\omega'})^{1/2}}{(1-e^{-8\pi
(M-\omega/2)\omega'})}\\ \times\frac{(1-e^{-8\pi
M\omega})^{1/2}(1-e^{-8\pi (M-\omega)\omega})^{1/2}e^{-8\pi
(M-\omega/2)\omega}}{(1-e^{-8\pi (M-\omega/2)\omega})^{2}} \,.
\end{split}
\end{equation}
And if $\omega\ll M$, it becomes
\begin{equation}
\label{eq:20}
\begin{split}
\tilde{N}(M,\omega)\approx \frac{1}{e^{8\pi
(M-\omega/2)\omega}-1}\approx e^{-8\pi (M-\omega/2)\omega}(M\gg 1)
\,,
\end{split}
\end{equation}
an analogous result as $\bar{N}(M,\omega)=(e^{8\pi
M\omega}-1)^{-1}$~\cite{a,b}, but with an $\omega^2$
correction~\cite{k}. In~\cite{k}, the authors treat
$e^{-8\pi (M-\omega/2)\omega}$ as a semiclassical emission rate,
but here it's $N(M,\omega)\approx e^{-4\pi M\omega}$ \footnote{The
estimate $N(M,\omega)\approx e^{-4\pi M\omega}$ agrees with
another computation in~\eqref{eq:63}.}that is a
part of the emission amplitude in~\eqref{eq:17}. If the absorption part in~\eqref{eq:14} is also included, some other factors may also be obtained,
for example a factor $\langle0_{M}\arrowvert\tilde{b}_{\omega}^{\dag}\tilde{b}_{\omega}\arrowvert0_M\rangle$ that
describes the process of the black hole emitting and re-absorbing. Even the factor $\bar{N}(M,\omega)$ can also be related to the process of the detector absorbing and re-emitting, in addition to the meaning of an expectation value of the particle number.

In fact, for some initial state $\arrowvert\phi\rangle$, by using of~\eqref{eq:14}, the expectation value of the
particle number for our model can be given by
\begin{equation}
\label{eq:21}
\begin{split}
\langle d_{\omega}^{\dag}d_{\omega}\rangle &= \langle\phi\arrowvert
U_{B,\psi,D}^{\dag}(d_{\omega}^{\dag}d_{\omega})U_{B,\psi,D}\arrowvert\phi\rangle=|g_{\omega}|^2\langle
b_{\omega}^{\dag}b_{\omega}\rangle \,.
\end{split}
\end{equation}
Furthermore, we can also calculate the correlation function
\begin{equation}
\label{eq:22}
\begin{split}
\langle(d_{\omega}^{\dag}d_{\omega})(d_{\omega'}^{\dag}d_{\omega'})\rangle=|g_{\omega}|^4\langle(b_{\omega}^{\dag}b_{\omega})(b_{\omega'}^{\dag}b_{\omega'})\rangle
=|g_{\omega}|^4\sum_{\alpha}\langle(b_{\omega}^{\dag}b_{\omega})\arrowvert\Phi_{\alpha}\rangle\langle\Phi_{\alpha}\arrowvert(b_{\omega'}^{\dag}b_{\omega'})\rangle
\,,
\end{split}
\end{equation}
where a completeness relation in~\eqref{eq:29} has been used. Since our
model is proposed in a unitary manner, then all the possible intermediate states will contribute
significantly, with some correlations being preserved among the radiations in the detector\footnote{The
correlations can not be completely preserved due to quantum
decoherence, but there are indeed some residual correlations, as
shown in~\eqref{eq:57}.}. These in fact imply that a distant exterior observer can acquire the information of the
radiations only by performing local
measurements on the detector. A general measurement is given by
\begin{equation}
\label{eq:23}
\begin{split}
\langle\phi\arrowvert U_{B,\psi,D}^{-1}\mathcal{O}_{D}U_{B,\psi,D}\arrowvert\phi\rangle &= tr_{D}(\mathcal{O}_{D}\rho_{D}) \,,
\end{split}
\end{equation}
for a detector's observable $\mathcal{O}_{D}$ that corresponds to $\mathcal{O}_{ext}$. The reduced density operator is
\begin{equation}
\label{eq:24}
\begin{split}
\rho_{D} &=
tr_{B,\psi}(U_{B,\psi,D}\arrowvert\phi\rangle\langle\phi\arrowvert
U_{B,\psi,D}^{-1}) \,,
\end{split}
\end{equation}
which leads to a super-operator evolution~\cite{i}
for the detector. And more discussions will be shown in section~\ref{sec:mix}.

\subsection{Some General Features of The Interactions }
\label{sec:feature}

Consider the vacuum ($\arrowvert0_M\rangle_{\psi}$) expectation
value of the full interaction term
\begin{equation}
\label{eq:25}
\begin{split}
\langle e^{-i\int d^4x(B+D)\psi}\rangle &= 1-i\langle\int
d^4x(B+D)\psi\rangle-\frac{1}{2}\langle[\int
d^4x(B+D)\psi]^2\rangle+\cdots \,,
\end{split}
\end{equation}
with $B$ and $D$ denoted as the black hole and the detector respectively. For an in-falling observer equipped with a detector, according to effective field theory this vacuum expectation value is the ordinary one without mode split. In other words, the field is expanded in terms of $a_{\omega}$ modes, so the detector will also receive radiations of $a_{\omega}$ modes. In this case, $B$ actually stands for the perturbation of a flat space-time, i.e. the gravitational perturbation. The first order term in~\eqref{eq:25} vanishes obviously, while the second and higher orders give the exchanges of energy among the components of the entire system. In the view of a static observer, the second order contains the following four processes
\begin{equation}
\label{eq:26}
\begin{split}
\langle[\int_{int} d^4xB\psi]^2\rangle \qquad\langle[\int_{ext}
d^4xD\psi]^2\rangle \\ \langle\int_{int} d^4xB\psi\int_{ext}
d^4yD\psi\rangle \qquad\langle\int_{ext} d^4xD\psi\int_{int} d^4yB\psi\rangle
\,,
\end{split}
\end{equation}
which still describe the energy exchanges or interactions between the components. The first two terms give the self-interactions of the black
hole and the detector themselves, while the last two describe the
interactions, or more exactly, nonlocal correlations between them. Since the space-time has been separated
into two causally disconnected regions, the first two self-interactions are well described in the framework of local effective field theory; while the last two with independent interaction terms can give non-trivial results only through the entanglement between the $b_{\omega}$ and $\tilde{b}_{\omega}$ modes implicit in the vacuum state $\arrowvert0_M\rangle$. The locality makes a static
exterior detector always receive radiations only in terms of $b_{\omega}$ modes; while for a static detector in the interior of the black hole but still far from the singularity at $r=0$, it will receive radiations only in terms of $\tilde{b}_{\omega}$ modes.

All of those terms in~\eqref{eq:26} involve the scalar field's propagator
denoted formally by $\langle\psi^2\rangle$. For an in-falling observer, it is an ordinary propagator, while for a static observer, it will depend on the relevant
modes due to the black hole's causal structure. For instance, the self-interaction terms make use of only $b_{\omega}$ or $\tilde{b}_{\omega}$ modes; while the correlations between the black hole and the detector should make use of both the $b_{\omega}$ and $\tilde{b}_{\omega}$ modes. Recalling the quantity $N(M,\omega)$ defined in~\eqref{eq:16}, except for the little difference of the vacua, it is just part of the propagator with contributions from both the $b_{\omega}$ and $\tilde{b}_{\omega}$ modes. In addition to the combination
in~\eqref{eq:14}, there is also another one, in which $\mathcal {H}_{B,\psi}$ is expanded as
$\sum_{\omega}[\mathbf{V}_{\omega}(t)\tilde{b}_{\omega}^{\dag}+h.c.]$
with $\mathbf{V}_{\omega}(t)\sim \int
d^3x\mathbf{V}(t)\tilde{u}^{*}_{\omega}$, while $H_{\psi,D}$ is $\sum_{\omega}[g_{\omega}(t)b_{\omega}
d_{\omega}+h.c.]$ with $g_{\omega}(t)d_{\omega}\sim \tilde{g}\int d^3x\phi_D(t)u_{\omega}$. As a consequence, another transition $\mathbf{V}_{\omega}(t)\langle\tilde{b}_{\omega}^{\dag}b_{\omega}^{\dag}\rangle
d_{\omega}^{\dag}$ for the evaporation can also be obtained, still with a contribution from the field propagator. At a first glance, it seems to be impossible to have expressions for $\mathbf{V}_{\omega}(t)$ and $d_{\omega}$ different from those used in the model of section~\ref{sec:interaction}. This is indeed possible if the black hole and the detector have their own mode expansions, so that four combinations can be constructed for each of the interaction terms $H_{B,\psi}$ and $H_{\psi,D}$.

In a view of evolution in the Heisenberg picture, there should be a following sequence for the field propagator
\begin{equation}
\label{eq:26a}
\begin{split}
\langle\psi_{a}^2\rangle\stackrel{formation}{\longrightarrow}\langle\psi_{b,\tilde{b}}^2\rangle\stackrel{evaporation}{\longrightarrow}\langle\psi_{a}^2\rangle
\,.
\end{split}
\end{equation}
To some extent, this sequence for the static observer demonstrates the unitary property of the evolution for both the formation and evaporation of the black hole, if no black hole is present at both the starting point and ending point of the evolution. If a local quantum measurement is performed on the scalar field as in the Hawking's arguments, the sequence in~\eqref{eq:26a} will be broken since a quantum measurement can lead to some non-unitary super-operator evolution. This can also be roughly explained as follows. According to quantum mechanics, an observable $\mathcal {O}_{ext}$ can be measured by means of a measurement apparatus via a coupling $\lambda\mathcal {O}_{ext}T_{app}$~\cite{i}, with $T_{app}$ a corresponding operator of the apparatus, while the back-reaction of the scalar field on the black hole can still be described by $H_{B,\psi}$. Then the vacuum expectation value of the full evolution operator is
\begin{equation}
\label{eq:26b}
\begin{split}
\langle\exp\{-i\int_{int} d^4xB\psi-it\lambda\mathcal {O}_{ext}T_{app}\}\rangle \,,
\end{split}
\end{equation}
where the two terms can not be combined to give a scalar field's propagator as in QFT, thus breaking the the sequence in~\eqref{eq:26a}. This is also the difference between our effective field coupling $H_{\psi,D}$ and the quantum measurement coupling $\lambda\mathcal {O}_{ext}T_{app}$.

In the semiclassical treatment, the
back-reaction of the scalar field on the black hole is treated by means of a semiclassical Einstein's equation, with one side the classical Einstein tensor,
while the other one an expectation value $\langle T_{\mu\nu}\rangle$. According to the extended BHC (ii') in section~\ref{sec:bhc}, there should be two semiclassical equations in both the interior and exterior of the event horizon respectively.
These two semiclassical equations can be well modelled by local interactions according to effective field theory.
The fundamental interaction is the coupling between the gravitational field or perturbation and the matter field via
\begin{equation}
\label{eq:27}
\begin{split}
\exp\{-i\int d^4xh_{\mu\nu}T^{\mu\nu}[\psi]\} \,,
\end{split}
\end{equation}
with $h_{\mu\nu}$ the gravitational perturbation. Because of the causal structure of the black hole, the energy-momentum tensor
$T^{\mu\nu}[\psi]$ is thus separated into two independent
parts, one part $T^{\mu\nu}[\psi_b]$ in terms of the $b_{\omega}$ modes, while the other one $T^{\mu\nu}[\psi_{\tilde{b}}]$ with the
$\tilde{b}_{\omega}$ modes. Analogously, the gravitational perturbation must also be split into two components that couple with the corresponding $T^{\mu\nu}[\psi_b]$ and $T^{\mu\nu}[\psi_{\tilde{b}}]$ respectively. Certainly, the proposed term
$H_{B,\psi}$ in our model is just an approximation to
the interior interaction. A somewhat precise treatment will
be given in section~\ref{sec:gra}.

\section{Dynamics of The Effective Field Model }
\label{sec:dynamics}

\subsection{The S-matrix Formula and Its Self-consistency}
\label{sec:S}
The S-matrix formula for a curved space QFT is more complicated than that for a flat space QFT because of the background dependence on the metric. An analogous but simple model is studied in appendix~\ref{sec:singular}, where the S-matrix formula is constructed in detail. In this subsection, we shall construct the S-matrix formula for our model following appendix~\ref{sec:singular}. The BHC (iii) indicates that the black hole's state $\arrowvert
M\rangle$ can be treated as a steady state, giving a stationary
space-time region. We can thus consider the transitions between
various states, which are well defined in two stationary regions with different black hole's masses, for example $M_1$ and $M_2$. In other words, an S-matrix formula between two arbitrary stationary regions can be constructed, at least for the restricted Hilbert space $\mathcal {H}_0$.

Recall $\mathcal {H}_0$ is spanned by the eigenstates of the free
Hamiltonian $H_0$ in~\eqref{eq:5}, with $H_{\psi}$ given by $H_{a}$ in~\eqref{eq:6b}. Thus near the infinite past $I^-$ in each stationary region, we have
\begin{equation}
\label{eq:28}
\begin{split}
H_{0i}\arrowvert\Phi_{i\alpha}\rangle &=
E_{\alpha}\arrowvert\Phi_{i\alpha}\rangle(E_{\alpha}=M_{i}+E_{i}),\qquad\arrowvert\Phi_{i\alpha}\rangle\equiv\arrowvert
M_{i}\rangle_{B}\arrowvert0_{M_{i}}\rangle_{\psi}\otimes\arrowvert
E_{i}\rangle_D,(i=1,2) \,,
\end{split}
\end{equation}
with $\alpha$ denoted as a collection of the quantum
numbers $(M,E)$. For those eigenstates, we also have the following relations
\begin{equation}
\label{eq:29}
\begin{split}
\langle\Phi_{j\beta}\arrowvert\Phi_{i\alpha}\rangle &=
\delta_{\beta\alpha}\delta_{ji},\qquad\sum_{\alpha}\arrowvert\Phi_{i\alpha}\rangle\langle\Phi_{i\alpha}\arrowvert=\mathbb{I}_i
\,,
\end{split}
\end{equation}
where the completeness relation~\eqref{eq:8} is extended by adding $\sum_{E}\arrowvert E\rangle_D\langle
E\arrowvert=\mathbb{I}_D$. Following appendix~\ref{sec:singular}, the S-operator can be constructed as
\begin{equation}
\label{eq:30}
\begin{split}
S(21) &= \Omega_2(+\infty)^{\dag}\Omega_1(-\infty) =
U_{21}(+\infty,-\infty) \,,
\end{split}
\end{equation}
where the evolution operator is defined as
\begin{equation}
\label{eq:31}
\begin{split}
U_{21}(t_2,t_1)\equiv \Omega_2(t_2)^{\dag}\Omega_1(t_1)=\exp(+i
H_{02}t_2)\exp\{-i H(t_2-t_1)\}\exp(-i H_{01}t_1) \,.
\end{split}
\end{equation}
Since the free field
Hamiltonian $H_{\psi}$ depends on the black hole's mass, we have $H_{01}\neq H_{02}$. Hence there is also a
singular initial condition $U_{21}(t_0,t_0)=e^{iH_{02}t_0}e^{-iH_{01}t_0}$ like~\eqref{eq:95}.
This singularity also occurs in Hawking's original arguments~\cite{a,b}, where near the two
stationary regions $I^-$ and $H^+\bigcup I^+$, the corresponding free Hamiltonians of the scalar field are
\begin{equation}
\label{eq:32}
\begin{split}
H_{a} &= \sum_{\omega}\omega a_{\omega}^{\dag}a_{\omega},\qquad
H_{b,\tilde{b}}=\sum_{\omega}\omega
b_{\omega}^{\dag}b_{\omega}+H_{\tilde{b}}\,.
\end{split}
\end{equation}
Since $H_{a}\neq H_{b,\tilde{b}}$, a singular evolution operator like the one in~\eqref{eq:31} can be also constructed. Therefore,
except for the possible singularity, the unitary is preserved formally\footnote{As analyzed in Appendix~\ref{sec:singular}, that singularity may be from the semiclassical property of curved space QFT, i.e. the classical metric. In fact, for our model, the black hole and the scalar field can be combined formally as
a single component, with the corresponding Hamiltonian $H_{B}$ and state $\arrowvert\phi_M\rangle_B$.
Then the state for the entire system can be expressed as
$\sum_{M+E=M_0}C(M)\arrowvert\phi_M\rangle_{B}\arrowvert
E\rangle_D$, and the relevant interaction can be constructed as
$\sum_{\omega}(\mathbf{V}_{\omega}d_{\omega}^{\dag}+h.c.)$,
leading to the correlation between the black hole (together with the scalar
field) and the detector. But this treatment is useless for our understanding of the black hole physics.}.

The evolution in~\eqref{eq:30} can be extended to a general one with (meta-)stable sequences for the states of the black hole and the detector
\begin{subequations}\label{eq:33}
\begin{align}
\label{eq:33:1} \arrowvert 0\rangle_B\rightarrow \arrowvert
M_1\rangle_B\rightarrow\cdots \rightarrow\arrowvert M_i\rangle_B
\rightarrow\arrowvert M_{i+1}\rangle_B\rightarrow \cdots \,,
\\
\label{eq:33:2} \arrowvert M_0\rangle_D\rightarrow \arrowvert
E_1\rangle_D\rightarrow\cdots \rightarrow\arrowvert E_i\rangle_D
\rightarrow\arrowvert E_{i+1}\rangle_D\rightarrow \cdots \,,
\end{align}
where the energy is roughly balanced between the black hole and detector, i.e. $M_i+E_i\approx M_0$ for every $i$. As for the scalar field,
the corresponding sequence of the vacua is
\begin{equation}
\label{eq:33:3} \arrowvert0_{0}\rangle_\psi\rightarrow
\arrowvert0_{M_1}\rangle_\psi\rightarrow\cdots
\rightarrow\arrowvert0_{M_i}\rangle_\psi
\rightarrow\arrowvert0_{M_{i+1}}\rangle_\psi\rightarrow \cdots \,.
\end{equation}
\end{subequations}
A complete sequence of the Cauchy surfaces along the sequences~\eqref{eq:33} may be chosen as
\begin{equation}
\label{eq:34}
\begin{split}
I^-_{0}\stackrel{1}{\longrightarrow} H^+_{1}\bigcup
I^+_{1}\stackrel{2}{\longrightarrow} H^+_{2}\bigcup
I^+_{2}\rightarrow\cdots \,,
\end{split}
\end{equation}
with the process 1 denoted as the formation of a black hole, i.e. the evolution in Hawking's arguments. The corresponding sequence of the free Hamiltonians is then given by\footnote{From this sequence of the free Hamiltonians and the subsequent evolution in~\eqref{eq:36}, it indicates that the full evolution must have been out of the space $\mathcal {H}_0$, since $\arrowvert0_M\rangle$ is not an eigenstate of $H_{b,\tilde{b}}$. Hence we should apply an extended space with $\mathcal {H}_0$ just as a subspace. Some more discussions will be shown in section~\ref{sec:ext}.}
\begin{equation}
\label{eq:35}
\begin{split}
(H^{BD}_{0}+H_{a_0})\stackrel{H_1}{\longrightarrow}
(H^{BD}_{0}+H_{b_1,\tilde{b}_1})\stackrel{H_2}{\longrightarrow}
(H^{BD}_{0}+H_{b_2,\tilde{b}_2})\rightarrow\cdots \,,
\end{split}
\end{equation}
with $H_{a}$ and $H_{b,\tilde{b}}$ given by~\eqref{eq:32}, and
the interaction terms $H_{B,\psi}$ and $H_{\psi,D}$ of our model are
contained in the full Hamiltonians $H_1$, $H_2$ and subsequent ones in~\eqref{eq:35}.

\begin{figure}[tbp]
\setlength{\unitlength}{1mm} \centering
\includegraphics[width=2.5in]{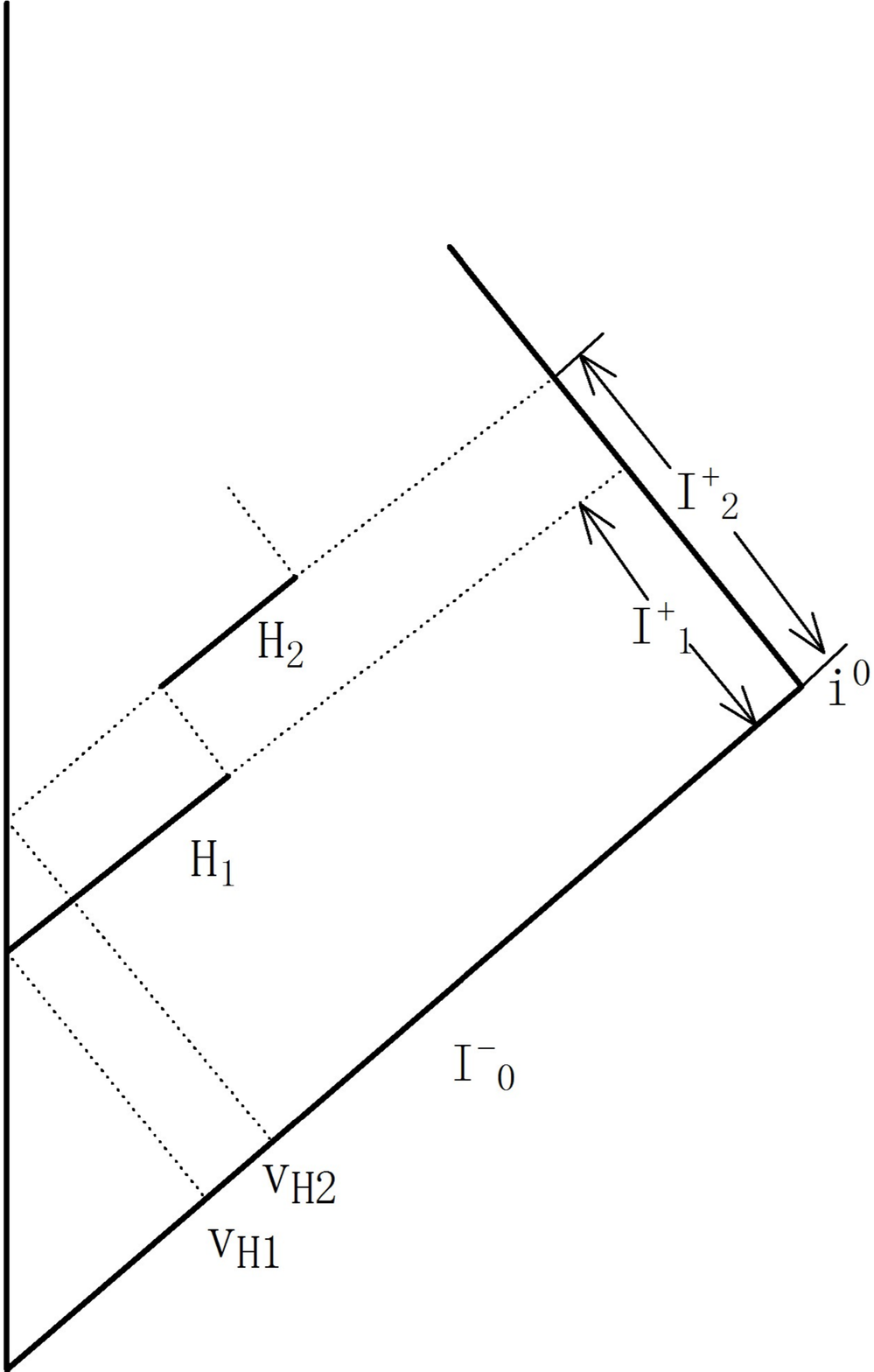}
\caption{\label{fig:1} The first two steps of a whole evolution, which is described in~\eqref{eq:36} with the Cauchy surfaces given in~\eqref{eq:34}.}
\end{figure}

\begin{figure}[tbp]
\setlength{\unitlength}{1mm} \centering
\includegraphics[width=3.0in]{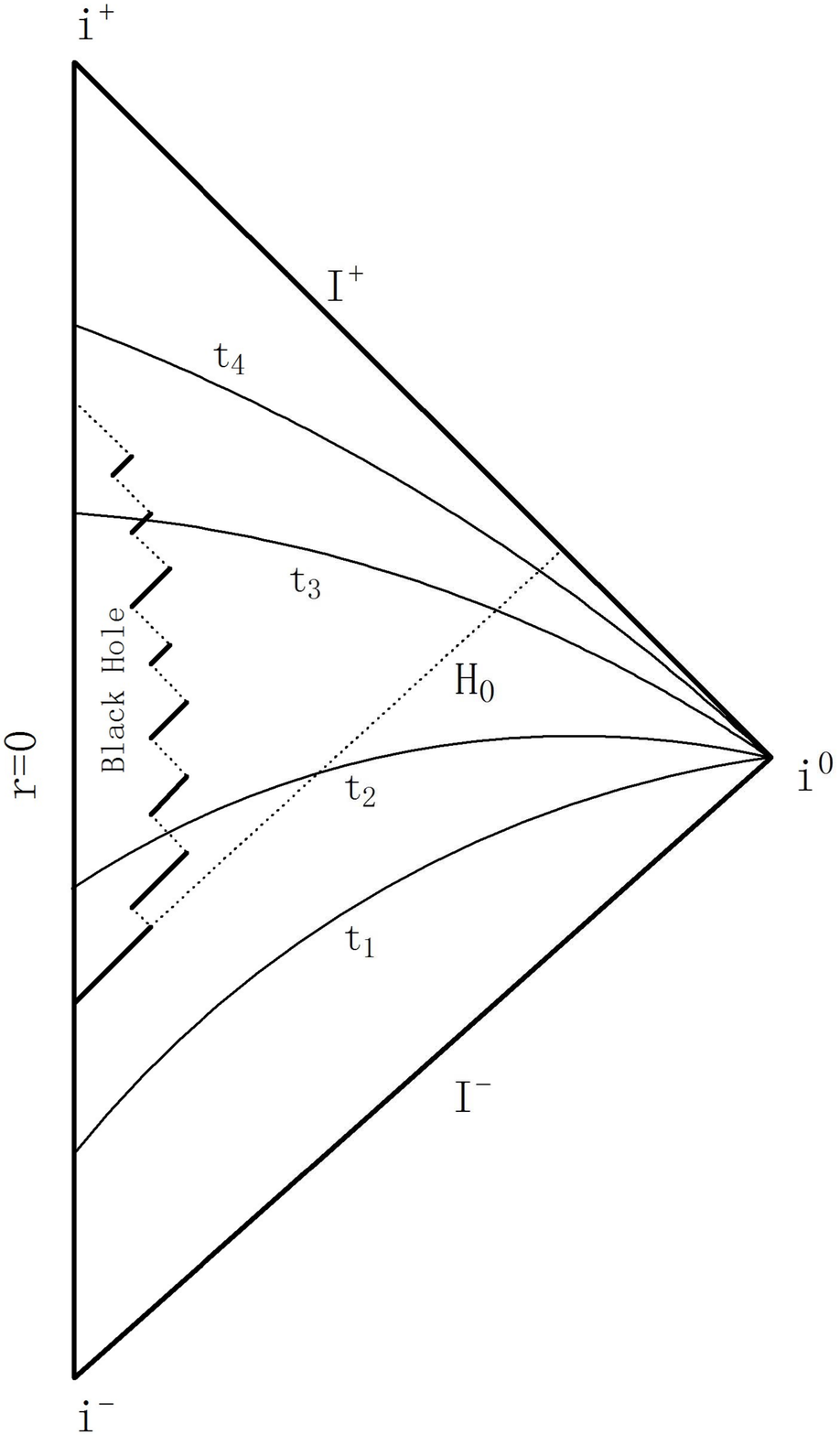}
\caption{\label{fig:2}Black hole evaporation: without extra matter
absorptions, the black hole will evaporate completely in the end,
leaving a space-time without any black hole.}
\end{figure}

The first two steps of the above sequences are
depicted in figure~\ref{fig:1}, and the required evolution operator can
be constructed formally as
\begin{equation}
\label{eq:36}
\begin{split}
U(t_2,t_1)U(t_1,t_0) = [e^{i(H^{BD}_{0}+H_{b_2,\tilde{b}_2})t_2}e^{-iH_2(t_2-t_1)}e^{-i(H^{BD}_{0}+H_{b_1,\tilde{b}_1})t_1}]\\
\times
[e^{-i(H^{BD}_{0}+H_{b_1,\tilde{b}_1})t_1}e^{-iH_1(t_1-t_0)}e^{-i(H^{BD}_{0}+H_{a_0})t_0}] \\= e^{i(H^{BD}_{0}+H_{b_2,\tilde{b}_2})t_2}\{e^{-iH_2(t_2-t_1)}e^{-iH_1(t_1-t_0)}\}e^{-i(H^{BD}_{0}+H_{a_0})t_0}\,.
\end{split}
\end{equation}
The term in the brace in the last line describes the processes of the formation and evaporation of a black hole, which is so complicated that we can not give an exact expression. However, a full Hamiltonian $H_{20}$ exists so that
\begin{equation}
\label{eq:36a}
\begin{split}
e^{-iH_2(t_2-t_1)}e^{-iH_1(t_1-t_0)}=e^{-iH_{20}(t_2-t_0)}\,,
\end{split}
\end{equation}
since there is always such a time parametrization satisfying
\begin{equation}
\label{eq:36b}
\begin{split}
H_{20}(t) = \left\{ \begin{array}{ll}
H_{1}(t)\qquad & \textrm{$t_0\leq t\leq t_1$}\\
H_{2}(t)\qquad & \textrm{$t_1\leq t\leq t_2$}
\end{array} \right.\,.
\end{split}
\end{equation}
Then \eqref{eq:36} will become
\begin{equation}
\label{eq:36c}
\begin{split}
U(t_2,t_1)U(t_1,t_0)= e^{i(H^{BD}_{0}+H_{b_2,\tilde{b}_2})t_2}e^{-iH_{20}(t_2-t_0)}e^{-i(H^{BD}_{0}+H_{a_0})t_0}=U(t_2,t_0)\,,
\end{split}
\end{equation}
which is the required associative relation for the evolution operator. In a functional form, this evolution can
also be described as
\begin{equation}
\label{eq:37}
\begin{split}
\langle\Psi(\Sigma_{t_2})\arrowvert\Psi(\Sigma_{t_1})\rangle &=
\int\mathcal {D}[B]\mathcal{D}[\psi]\mathcal{D}[D]\exp
 i\int_{t_1}^{t_2} d^4x\{\mathcal {L}_0[B,\psi,D]+(B+D)\psi\} \,,
\end{split}
\end{equation}
with $B$ and $D$ still denoted as the black hole and the detector,
respectively. Analogously, the complete S-operator $U(+\infty,-\infty)$ can be constructed step by step following the sequences~\eqref{eq:33}. Since those sequences are chosen arbitrarily, there may be one in which the mass of an already formed black hole continues decreasing so that the
black hole disappears in the end, as depicted in figure~\ref{fig:2}.
In this figure, the long dash line is the event horizon of the formed black hole at $r_{H_0}=2M_0$,
while the region surrounding by the zigzag line is the black hole interior, due to
decreasing of the event horizon in size during the evaporation.

There are also four typical t-slices (space-like Cauchy
surfaces) in figure~\ref{fig:2}, foliating the space-time. Similarly
as reference~\cite{c}, for the $t_2$ slice, because it
intersects the interior of the black hole, the state on it should be
expressed as
\begin{equation}
\label{eq:38}
\begin{split}
\arrowvert\Psi(\Sigma_{t_2})\rangle &= \sum_{i,j}\arrowvert
\phi_i(\Sigma_{int})\rangle_{B,\psi}\arrowvert\chi_j(\Sigma_{ext})\rangle_{\psi,D}
\,,
\end{split}
\end{equation}
with the sum of the indexes $i,j$ denoted as correlations between the interior and exterior degrees of freedom due to the interactions $H_{B,\psi}$ and $H_{\psi,D}$. Notice~\eqref{eq:38} is different from the one used in~\cite{c}, where the full state was given by a direct tensor product of the interior and exterior components, since there were no interactions that induce correlations between the interior and exterior modes in their article. Moreover, in the evaporation figure of~\cite{c}, the interior region of the black hole is depicted as if it suddenly disappeared just after the complete evaporation. While in figure~\ref{fig:2} the gradual changes of the event horizon is demonstrated apparently, that is, the interior region becomes exterior gradually during the evaporation. On the $t_3$ slice the evolution continues, but with less interior modes than those on the $t_2$ one. At last, on the $t_4$ slice the black hole disappears completely, so the interaction $H_{B,\psi}$ has stopped. And we will obtain a final pure state
$\arrowvert\Psi(\Sigma_{t_4})\rangle$, with the remaining
correlations stored among the scalar field, the weak gravitational field and the detector because of the
interaction $H_{\psi,D}$ and the one in~\eqref{eq:27}. The evaporation depicted in figure~\ref{fig:2} is just a particular one of all the possible complete
evolutions that can be accomplished via~\eqref{eq:36} or~\eqref{eq:37} step by step.
Therefore, we can conclude that the whole process of formation and evaporation of a black hole is
unitary as the entire system evolves.

\subsection{Discussions on The Paradoxes of Information Loss and AMPS's Firewall }
\label{sec:infor}

As shown in the above subsection, the black hole evaporation based on our model is
unitary, then how to understand Hawking's information loss
arguments? According to~\cite{a}, a distant exterior observer
should construct an operator like $\mathcal {O}_{ext}$, which can act
only on the space generated by $b_{\omega}^{\dag}$. Then, if we
calculate the expectation value like
\begin{equation}
\label{eq:39}
\begin{split}
\langle0_M\arrowvert\mathcal {O}_{ext}\arrowvert0_M\rangle \,,
\end{split}
\end{equation}
the $\tilde{b}_{\omega}$ modes cancel automatically, leading to a
mixed state. This can also be seen by performing a partial trace
over the $\tilde{b}_{\omega}$ modes in the initial density operator
$\arrowvert0_M\rangle\langle0_M\arrowvert$, giving a super-operator
evolution for the exterior $b_{\omega}$ modes. In the discussions below~\eqref{eq:f} we have shown that, in the framework of effective field theory with background dependence, local measurements will always lead to (correlation) information loss in the black hole background. However, in the evolution of the last subsection, there are no local measurements. The radiation detector is not a standard measurement apparatus, but only serves as an environment coupling with the scalar field according to effective field theory.

In reference~\cite{g}, a ``pull-back-push-forward'' strategy is proposed to identify the interior
with the exterior degrees of freedom. It's assumed there are two unitary transformations $\mathcal {U}$ and
$\mathcal {V}$ that can evolve an initial full operator $\mathcal {O}$ as
\begin{equation}
\label{eq:42}
\begin{split}
\mathcal {U}^{-1}\mathcal {O}\mathcal {U} &= \mathcal
{O}_{ext},\qquad\mathcal {V}^{-1}\mathcal {O}\mathcal {V}=\mathcal
{O}_{int} \,.
\end{split}
\end{equation}
The exterior operator can thus be related to the interior one simply
by \cite{g}
\begin{equation}
\label{eq:43}
\begin{split}
\mathcal {O}_{ext} &= (\mathcal {U}^{-1}\mathcal {V})\mathcal
{O}_{int}(\mathcal {V}^{-1}\mathcal {U}) \,.
\end{split}
\end{equation}
Now let's go into the Schr\"odinger picture. Since $\mathcal {O}$ is
a full operator, in order for~\eqref{eq:39} to contain only an
``ext'' (or ``int'' ) sector, in general, it should be assumed that
\begin{equation}
\label{eq:44}
\begin{split}
\mathcal {U}\arrowvert0_M\rangle &=
\arrowvert\phi\rangle_{ext},\qquad\mathcal
{V}\arrowvert0_M\rangle=\arrowvert\chi\rangle_{int} \,,
\end{split}
\end{equation}
then we will have
\begin{equation}
\label{eq:45}
\begin{split}
\langle0_M\arrowvert\mathcal {V}^{-1}\mathcal
{U}\arrowvert0_M\rangle &=
_{int}\langle\chi\arrowvert\phi\rangle_{ext} = 0 \,.
\end{split}
\end{equation}
This means the unitary transformation $\mathcal {V}^{-1}\mathcal {U}$
would be singular, which is impossible. This inconsistency can also be
seen as follows. Consider two non-commutative full operators
$\mathcal {O}_1$ and $\mathcal {O}_2$, then from~\eqref{eq:42} we
can get
\begin{equation}
\label{eq:46}
\begin{split}
[\mathcal {O}_{1ext},\mathcal {O}_{2int}] &= 0 \,,
\end{split}
\end{equation}
that is, one representation can be found so that two
non-commutative operators are commutative, which is also impossible.
Therefore, the unitary transformations $\mathcal {U}$ and $\mathcal
{V}$ utilized in the ``pull-back-push-forward'' strategy can not be
constructed consistently. In fact, the observables $\mathcal {O}_{ext}$ and $\mathcal {O}_{int}$ are independent, and the exterior and interior degrees
of freedom can be related to each other only through the
correlation implicit in~\eqref{eq:f}, or more exactly, the entanglement implicit in the vacuum $\arrowvert0_M\rangle$.

However, the information stored in the correlation~\eqref{eq:f} or the entanglement implicit in the vacuum $\arrowvert0_M\rangle$ cannot be physically acquired due to the black hole's causal structure. For this reason, a radiation detector has been introduced in our model to only receive the radiation.
Then where is the possible physical information in our model? Without any
measurement, the black hole and the detector are effectively correlated or
entangled via the entanglement implicit in $\arrowvert0_M\rangle$, as
given roughly by the transition in~\eqref{eq:17}. \emph{The possible physical information of our model may be stored in
this new generated correlation or entanglement between them}. In this way, the firewall paradox can be resolved, since any state in the space $\mathcal
{H}_0$ can be both an $a_{\omega}$-vacuum and a
$d_{\omega}^{\dag}d_{\omega}$ eigenstate, instead of the
$b_{\omega}^{\dag}b_{\omega}$ as argued in~\cite{d}\footnote{Since $[a_{\omega},b_{\omega'}]\neq 0$,
it is impossible to construct a state to be eigenstate of both the
$a_{\omega}$ and $b_{\omega}^{\dag}b_{\omega}$.}. This is because in our model, via the interaction $H_{\psi,D}$ the radiation has been transferred into
the detector expressed in terms of $d_{\omega}$ modes, with the scalar field only as an intermediate medium. Under this circumstance, the arguments of the
AMPS's firewall don't hold any longer. In the firewall paradox, only
three subsystems are present, the black hole together with the $\tilde{b}_{\omega}$ modes as a whole, the early radiation and the late one, so that entanglements would be shared by two
pairs: one pair is the early and late radiations, while the
other one is the late radiation and $\tilde{b}_{\omega}$ modes behind the event horizon,
violating the monogamy of entanglement. In our model, there are \emph{four subsystems} effectively, the black hole, the detector or radiation in terms of $d_{\omega}$ modes,
and the splitting $b_{\omega}$ and $\tilde{b}_{\omega}$ modes of the scalar field. The radiation is transferred between the black hole and the detector($d_{\omega}$ modes), as indicated in~\eqref{eq:17}. This generates a new entanglement between them, replacing the entanglement in the firewall paradox between the early and late radiations in terms of $b_{\omega}$ modes. During the energy transfer, the vacuum state $\arrowvert0_{M}\rangle$ may be changed into other states, for example another vacuum
$\arrowvert0_{M'}\rangle$ with a different mass, but the entanglement implicit in
them almost remains, as will be shown in
section~\ref{sec:entan}. In a word, there are two different entanglements among two pairs of subsystems. \emph{This does not violate
the monogamy of entanglement}, and the firewall paradox is thus resolved.

The generation of entanglement through an old one is different from
the ``transfer of entanglement'' in reference~\cite{g}, where the
author says that ``as time evolves entanglements shift from the
near horizon region to the evaporation products of the black hole".
That is, the entanglements between the $b_{\omega}$ and
$\tilde{b}_{\omega}$ modes are all transferred into the whole exterior
radiation space, leading to a sort of ``conservation of entanglement''~\cite{f}. This is impossible by noting that \emph{the
energy of the radiations should not only come from the near horizon region with state $\arrowvert0_M\rangle$, but also come
from the black hole itself with state $\arrowvert M\rangle$}, as shown in our model. It is always believed the expectation value of the energy-momentum tensor $\langle0_M\arrowvert T_{\mu\nu}\arrowvert0_M\rangle$ provides the total energy extracted from the black hole, via the semiclassical Einstein's equation. But this is not the case. From the Hawking's evolution given by~\eqref{eq:32}, one can see that the initial average energy of the scalar field $\langle0_M\arrowvert H_a\arrowvert0_M\rangle$ is exactly 0, while the final one $\langle0_M\arrowvert H_{b,\tilde{b}}\arrowvert0_M\rangle$ is of the order $M^{-2}$. This energy difference or vacuum energy should not be treated as the total evaporated energy since a contribution from the interior $\tilde{b}_{\omega}$ modes is also included. Moreover, for large $M$ this contribution is negligible and can be regarded to be \emph{the work done by the black hole on the scalar field due to the causal structure}, just like the case in which gravitational potential energy is transported into the matter via the work done by the gravity. But for small $M$, that energy difference or vacuum energy will become divergent and dominate, implying the need of a quantum gravity theory\footnote{These can also be explained by the back-reaction of the matter on the black hole. For a macro black hole with a large mass $M$, the energy of the matter is small compared with $M$, so the back-reaction or the perturbation of the black hole is small, and effective field theory is valid. While for a micro black hole with a small mass $m$, the perturbation may be large when the contributed energy from the matter exceed $m$, then a quantum gravity theory will be needed.}.

In our model, however, the energy transfer between the black hole and the detector is accomplished via the quantum transitions given roughly by~\eqref{eq:17}. We can still divide the radiations into an early and a late part, as a whole, these two parts should be
entangled. The early part has already been transferred into the detector,
while the late part is still stored in the black hole. Since the
detector is entangled with the black hole via the new generated
entanglement, the early and late parts of the radiations are thus
entangled, too. This can be simply expressed as
\begin{equation}
\label{eq:47a}
\begin{split}
\sum_{M}C(M)\arrowvert
M\rangle_B\arrowvert0\rangle_D\longrightarrow\sum_{M\omega_1\omega_2}(\gamma_{M\omega_1\omega_2}\arrowvert
M-\omega_1-\omega_2\rangle\arrowvert\omega_2\rangle)_B\arrowvert\omega_1\rangle_D \,,
\end{split}
\end{equation}
where the late radiation state $\arrowvert\omega_2\rangle$ can be treated
as a part of the black hole's state before it is transferred into the
detector. This means that \emph{the correlations among the radiations are mainly provided by those correlations within the states of the black
hole, together with the nonlocal correlations between the black hole and the detector}. In other words, the physical information of our model is mainly stored in the correlations or entanglements between the black hole and the detector during the evaporation. When the black hole evaporates completely, the information can still be stored in the correlations among the detector, a weak gravitational field and the scalar field. Whether this information can be acquired will be discussed in the next subsection.

In conclusion, the physically acquirable information is not stored in
the entanglement implicit in the vacuum $\arrowvert0_M\rangle$, but may be in
a new entanglement between the black hole and the detector generated
via the one implicit in $\arrowvert0_M\rangle$. Without
any local measurement, this information can not be lost for ever since the entire system evolves in a unitary manner, as shown in section~\ref{sec:S}.
Moreover, the new generated entanglement does not violate the monogamy of
entanglement, thus no AMPS's firewall could emerge.

\subsection{Quantum Decoherence Due to Measurement on The Detector}
\label{sec:mix}

As shown above, the evolution of the entire system is
unitary, as long as there are no local measurements. Under this
circumstance, the information won't be lost. In other words, if there is any
information loss, it must be attributed to some local
measurement, for instance, the measurement~\eqref{eq:23} on the detector space or Hawking's measurement~\eqref{eq:39} on the exterior modes. These two measurements are both local in the sense that they are performed only on some component of a bigger system.
These partial measurements can be described well by the so called
positive operator-valued measure (POVM)~\cite{i} approach, leading
to a super-operator evolution for the measured component, just like~\eqref{eq:24}. Since any measurement outcome must be definite, the
original quantum coherence among the components of the bigger
system disappears after the measurement, i.e. the so called \emph{quantum
decoherence}. The measurement~\eqref{eq:39} used by Hawking can always lead to information loss due to the black hole's causal structure, as shown in section~\ref{sec:bhc}. In this subsection, we consider the measurement~\eqref{eq:23} on the detector and show the information loss
due to quantum decoherence during the black hole evaporation.

As shown previously, the black hole and the detector are effectively entangled, so we can
consider a black hole-detector system by performing a partial trace
over the scalar field space. The coupling between them can be
approximated by an linear operator\footnote{Since we have performed
a partial trace over the scalar field space, the operator $L_{B,D}$ can not
be unitary, thus leading to some information loss. But the lost information
is all about the scalar field and is not the relevant physical
information of our model, as analyzed in the last subsection.}
\begin{equation}
\label{eq:48}
\begin{split}
L_{B,D} &=
1-\sum_{\omega}(\lambda_{\omega}\mathbf{V}_{\omega}d_{\omega}^{\dag}+h.c.)+\cdots
\,,
\end{split}
\end{equation}
where the lowest order term in~\eqref{eq:17} has been used, and
the coupling constant is given by
\begin{equation}
\label{eq:49}
\begin{split}
\lambda_{\omega}\simeq N(M,\omega)g_{\omega} \,.
\end{split}
\end{equation}
Except for the non-unitary, this coupling resembles the
amplitude-damping channel~\cite{i}, a schematic model of the decay
of an excited state of an atom due to spontaneous emission of
photons. Assuming that the initial state is
$\rho_B(0)\otimes\arrowvert0\rangle_D\langle0\arrowvert$, and that
there is no extra matter absorbed by the black hole, the
evolution of $\rho_B$ in the interaction picture can be described by
a first order Lindblad's equation~\cite{i}
\begin{equation}
\label{eq:50}
\begin{split}
\dot{\rho_B}\approx
\sum_{\omega}|\lambda_{\omega}|^2(\mathbf{V}_{\omega}\rho_B\mathbf{V}_{\omega}^{\dag}
-\frac{1}{2}\mathbf{V}_{\omega}^{\dag}\mathbf{V}_{\omega}\rho_B-\frac{1}{2}\rho_B\mathbf{V}_{\omega}^{\dag}\mathbf{V}_{\omega})
\,.
\end{split}
\end{equation}
We can further make an operator factorization as
$\mathbf{V}_{\omega}=V_{M,\omega}\mathbf{c}$, with $\mathbf{c}$ an
annihilator-like operator acting on the states of the black hole.
Then following~\cite{i}, we have
\begin{equation}
\label{eq:51}
\begin{split}
\langle\mathbf{c}_{\omega}^{\dag}\mathbf{c}_{\omega}(t)\rangle\sim
e^{-|\lambda_{\omega}V_{M,\omega}|^2t} \,,
\end{split}
\end{equation}
which is the familiar exponential law for a decay, giving an estimate of total decay rate
$\sum_{\omega}|\lambda_{\omega}V_{M,\omega}|^2$. That is,
in the view of a black hole alone, the evaporation is like \emph{a
non-unitary decay process}. This can be explained by quantum decoherence. Actually, to obtain the Lindblad's
equation~\eqref{eq:50}, we have performed a partial trace over the
detector space
\begin{equation}
\label{eq:52}
\begin{split}
\dot{\rho_B} &=
tr_{D}[L_{B,D}(0)\rho_B(0)\otimes\arrowvert0\rangle_D\langle0\arrowvert
L_{B,D}^{\dag}(0)] \,,
\end{split}
\end{equation}
with the first order Kraus representation operator given
by\footnote{The general Kraus representation of the
super-operator evolution is
$\$(\rho_B)=\sum_{\mu}M_{\mu}\rho_BM_{\mu}^{\dag}$, where the
operator $M_{\mu}=_{\psi,D}\langle\mu\arrowvert
U_{B,\psi,D}\arrowvert0_M\rangle_{\psi}\arrowvert0\rangle_D$ with the
$\arrowvert\mu\rangle_{\psi,D}$ denoted as the state of both the scalar field and the
detector.}
\begin{equation}
\label{eq:53}
\begin{split}
M_{\omega} &= _D\langle\omega\arrowvert
L_{B,D}(0)\arrowvert0\rangle_D=-\lambda_{\omega}\mathbf{V}_{\omega}
\,.
\end{split}
\end{equation}
Roughly speaking, during the black hole evaporation, we keep
performing partial measurements on the detector to determine the
back-reactions on the black hole approximatively, i.e. the Markovian approximation~\cite{i}.
As a consequence, the information will be lost due to quantum decoherence.

Now, let's see whether the information is completely lost, in the sense
of Hawking's completely thermal spectrum~\cite{j}
\begin{equation}
\label{eq:54}
\begin{split}
\langle0_M\arrowvert
b_{\omega_1}^{\dag}b_{\omega_1}b_{\omega_2}^{\dag}b_{\omega_2}\arrowvert0_M\rangle
&= \langle0_M\arrowvert
b_{\omega_1}^{\dag}b_{\omega_1}\arrowvert0_M\rangle\langle0_M\arrowvert
b_{\omega_2}^{\dag}b_{\omega_2}\arrowvert0_M\rangle \,.
\end{split}
\end{equation}
In our model, the correlation is given in~\eqref{eq:22}. By using of
the approximation in~\eqref{eq:48}, the general state of
the black hole-detector system can be expressed as
\begin{equation}
\label{eq:55}
\begin{split}
\arrowvert\phi\rangle = \epsilon\arrowvert
M\rangle_B\arrowvert0\rangle_D+\sum_{\omega}\alpha_{\omega}\arrowvert
M-\omega\rangle_B|\omega\rangle_D
+\sum_{\omega_1\omega_2}\beta_{\omega_1\omega_2}\arrowvert
M-\omega_1-\omega_2\rangle_B\arrowvert\omega_1,\omega_2\rangle_D+\cdots
\,.
\end{split}
\end{equation}
Then the reduced density operator $\rho_D$ in~\eqref{eq:24} will be expressed as
\begin{equation}
\label{eq:56}
\begin{split}
\rho_D &=
|\epsilon|^2\arrowvert0\rangle_D\langle0\arrowvert+\sum_{\omega}|\alpha_{\omega}|^2\arrowvert\omega\rangle_D\langle\omega\arrowvert
+\sum_{\omega_1\omega_2}|\beta_{\omega_1\omega_2}|^2\arrowvert\omega_1,\omega_2\rangle_D\langle\omega_1,\omega_2\arrowvert+\cdots
\,,
\end{split}
\end{equation}
i.e. a statistical ensemble. After some calculations, we then get
\begin{equation}
\label{eq:57}
\begin{split}
tr(\rho_Dd_{\omega_1}^{\dag}d_{\omega_1}d_{\omega_2}^{\dag}d_{\omega_2})\neq
tr(\rho_Dd_{\omega_1}^{\dag}d_{\omega_1})tr(\rho_Dd_{\omega_2}^{\dag}d_{\omega_2})
\,,
\end{split}
\end{equation}
because of the probability relation $P(\omega_1,\omega_2)\neq
P(\omega_1)P(\omega_2)$. This implies that the radiations in the
detector are not completely independent, instead some correlations or entanglements are remaining. In this sense,
\emph{the information is not completely lost}. This can
simply be explained by noting the intermediate states in~\eqref{eq:22}. Although the information is lost partially, the lost information cannot be recovered as long as the measurement is performed during the black hole evaporation. This is because, according to the analysis below~\eqref{eq:f}, the correlation information between the black hole and the detector cannot be fully acquired due to the black hole's causal structure. In this case, \emph{quantum measurement should be performed after the black hole evaporate completely}, otherwise (correlation) information will be lost inevitably due to the causal disconnectedness\footnote{This emphasized condition is proposed only in the framework of effective field theory, not for a complete quantum gravity theory, as indicated by the analysis below~\eqref{eq:f}. Notice further that in Hawking's arguments, local measurement such as $\langle0_M\arrowvert\mathcal {O}_{ext}\arrowvert0_M\rangle$ serves as the source term of the semiclassical Einstein equation. As a consequence, information is always lost during the evaporation process in this way, in particular the correlation information. However, in our model black hole can evaporate completely in a unitary manner, so the emphasized condition can avoid the correlation information loss effectively. }.

In conclusion, by performing some local quantum
measurements on the detector during the black hole evaporation, we will obtain mixed states
inevitably because of the black hole's causal structure and quantum decoherence. As a result, the
information must be lost, but only partly, giving a
\emph{non-thermal} property of the radiations. However, after the black hole evaporate completely, local measurement can lead to information loss only due to ordinary quantum decoherence\footnote{Recall that a quantum measurement only gives probabilities, with the phase information lost. For an entangled state such as an EPR pairs, correlation information can be acquired by comparing measurement outcomes through a classical channel. Thus, without black holes, only phase information can be lost due to quantum decoherence.}.

\subsection{Extensions Including Excited States of The Scalar Field}
\label{sec:ext}

In the space $\mathcal {H}_0$ given by~\eqref{eq:2}, the scalar field is assumed to be always in its (near horizon) vacuum state depending on some particular black hole's mass. In this subsection, we shall make some extensions to include excited states. But first, let's consider another extension about the dependence on the black hole's mass, which has been ignored in the previous analysis. For the vacuum state $\arrowvert0_M\rangle$ in~\eqref{eq:4}, in addition to the explicit
dependent factor $e^{-4\pi M\omega}$, the creators $b_{\omega}^{\dag}$ and
$\tilde{b}_{\omega}^{\dag}$ can also depend on the black hole's mass,
as indicated by the sequence~\eqref{eq:35}. In the calculations of the factors
$\langle0_{M-\omega}\arrowvert\tilde{b}_{\omega}\arrowvert0_M\rangle$
in~\eqref{eq:12} and $N(M,\omega)$ in~\eqref{eq:16}, the mass dependence of the creators and annihilators has been neglected, by assuming these operators belong to one Fock space with some fixed black hole's mass. Then one may ask whether the results would become very different, if including the mass dependence of the creators and annihilators.

As an example, let's consider a general factor $\langle0_{M_1}\arrowvert\tilde{b}^0_{\omega}\arrowvert0_{M_0}\rangle$ with
\begin{equation}
\label{eq:58}
\begin{split}
a^0_{\omega}\arrowvert0_{M_0}\rangle &=
a^1_{\omega}\arrowvert0_{M_1}\rangle = 0 \,,
\end{split}
\end{equation}
with the indexes 0 and 1 denoted as the mass
dependence of $M_0$ and $M_1$ respectively. The annihilators
$a^0_{\omega}$ and $a^1_{\omega}$ are related by a Bogoliubov
transformation, and there is also a relation between the two vacua
\begin{equation}
\label{eq:59}
\begin{split}
\arrowvert0_{M_1}\rangle\sim
\exp\{\gamma_{\omega\omega'}(a^0_{\omega})^{\dag}(a^0_{\omega'})^{\dag}\}\arrowvert0_{M_0}\rangle
\,.
\end{split}
\end{equation}
Then we have
\begin{equation}
\label{eq:60}
\begin{split}
\langle0_{M_1}\arrowvert\tilde{b}^0_{\omega}\arrowvert0_{M_0}\rangle\sim
\langle0_{M_0}\arrowvert\exp\{\gamma_{\omega\omega'}a^0_{\omega}a^0_{\omega'}\}\tilde{b}^0_{\omega}\arrowvert0_{M_0}\rangle
= 0 \,,
\end{split}
\end{equation}
where another Bogoliubov transformation $\tilde{b}^0\sim \alpha
a^0+\beta (a^0)^{\dag}$ has been used. From~\eqref{eq:60} we see that, as long as the initial
state is some (near horizon) vacuum, particles of both the two modes must be created or annihilated in pairs
(generally in even numbers), leaving the vacuum almost unaltered. This is also the reason for the non-vanishing of the factor
$N(M,\omega)$ which contains annihilators of both the two modes. By including the mass dependence of the creators and annihilators, it
may be extended to a non-diagonal factor $N(M,\omega,\omega')$, which leads to a general
correlation $\sum_{\omega \omega'}f_{\omega \omega'}\arrowvert
M-\omega\rangle_B\arrowvert E+\omega'\rangle_D$ between the black hole and the detector.

Now, let's consider the excited states of the scalar field. As shown in section~\ref{sec:S}, the space $\mathcal {H}_0$ is not enough for a full unitary evolution, that is, excited states can be obtained during the evolution. In the evolution operator~\eqref{eq:36}, the completeness relations in terms of $b_{\omega}$ and $\tilde{b}_{\omega}$ modes instead of $a_{\omega}$ modes, should be inserted. It thus seems that the evolution may be described in a direct product Hilbert space
$\mathcal {H}_b\otimes \mathcal {H}_{\tilde{b}}$, with the vacuum $\arrowvert0,\tilde{0}\rangle$. Then the creation and
annihilation of particles of the $b_{\omega}$ and $\tilde{b}_{\omega}$ modes
may be un-related, behaving like two independent dynamical
systems. However, given a single system with an initial state $\arrowvert0_M\rangle$, it seems to be impossible to
obtain two completely independent subsystems under a unitary evolution. This can be roughly explained as follows. With the interactions of our model being turned on, one would obtain some states which can be expressed as, for example $(b)^m(b^{\dag})^n(\tilde{b})^p(\tilde{b}^{\dag})^q\arrowvert0_M\rangle$, i.e. with finite creators and annihilators acting on the initial vacuum. They can be derived from terms like $(a^{\dag})^l\arrowvert0_M\rangle$, with a relation $l=m+n+p+q$ due to the Bogoliubov transformations. This relation should be preserved under the unitary evolution, meaning that the completeness relation for the Hilbert space $\mathcal {H}_a$ of the $a_{\omega}$ modes can still be used. When expressed in terms of the $b_{\omega},\tilde{b}_{\omega}$ modes, it is identified with the completeness relation for the direct product space $\mathcal {H}_b\otimes \mathcal {H}_{\tilde{b}}$. If further $l=2k$, i.e. an even number, then from~\eqref{eq:59} the scalar field can be in some near horizon vacuum state, but only with a small probability for large $M$, indicated by the estimate in~\eqref{eq:20}\footnote{The existence for the case of $l=2k+1$ implies that we should work in a larger Hilbert space, with $\mathcal {H}_0$ only as a subspace.}. Moreover, the relation~\eqref{eq:59} implies that the creators are required to be combined in some coherent manner to obtain another vacuum $\arrowvert0_{M_1}\rangle$ from an initial one $\arrowvert0_{M_0}\rangle$, which can not be achieved simply by using of only a perturbation method. Thus, in terms of only perturbations, states excited relative to some chosen vacuum $\arrowvert0_{M}\rangle$ could be obtained as long as the radiations have not been absorbed by some other systems.

In the lowest order, if the initial state is the vacuum $\arrowvert0_M\rangle$, the states of the scalar field in each subsequent stationary region can be assumed to be expressed as\footnote{It should be stressed that the state~\eqref{eq:61} is actually not exact and even not present, it is proposed only for convenience of calculation. }
\begin{equation}
\label{eq:61}
\begin{split}
\arrowvert\phi\rangle &=
\alpha\arrowvert0,\tilde{0}\rangle+\beta\arrowvert1,\tilde{1}\rangle+\cdots+\gamma\arrowvert
n,\tilde{n}\rangle+\cdots \,,
\end{split}
\end{equation}
with all the possible $\omega$ labels neglected for simplicity. This form is possible only when the radiations emitted from the back hole are almost absorbed by the detector. As a
coherence state, the near horizon vacuum is just a
particular one of them. Instead of~\eqref{eq:8}, another approximate completeness relation can be utilized for the scalar field space
\begin{equation}
\label{eq:62}
\begin{split}
\mathbb{I} =
\arrowvert0,\tilde{0}\rangle\langle0,\tilde{0}\arrowvert+\sum_{\omega}\arrowvert\omega,\tilde{\omega}\rangle\langle\omega,\tilde{\omega}\arrowvert+
\sum_{\omega_1\omega_2}\arrowvert\omega_1\omega_2,\tilde{\omega_1}\tilde{\omega_2}\rangle\langle\omega_1\omega_2,\tilde{\omega_1}\tilde{\omega_2}\arrowvert+\cdots \,.
\end{split}
\end{equation}
Then some transition rates can be calculated, where the mass dependence of the creators and annihilators can still be neglected for simplicity.
For example, the transition rate for $\arrowvert0_M\rangle$ to
$\arrowvert0,\tilde{0}\rangle$ is given by
\begin{equation}
\label{eq:63}
\begin{split}
|\lambda^{0}_{\omega}V_{M,\omega}|^2=|\langle0,\tilde{0}\arrowvert
b_{\omega}\tilde{b}_{\omega}\arrowvert0_M\rangle
g_{\omega}V_{M,\omega}|^2\simeq |g_{\omega}V_{M,\omega}|^2e^{-8\pi
M\omega} \,,
\end{split}
\end{equation}
where we have used an approximation
\begin{equation}
\label{eq:64}
\begin{split}
\prod_\omega(1-e^{-8\pi M\omega}) &= e^{-\frac{1}{96M}}\simeq 1,
(M\gg 1) \,.
\end{split}
\end{equation}
This gives a black hole mass decreasing rate
\begin{equation}
\label{eq:65}
\begin{split}
\frac{1}{2\pi}\int_{0}^{\infty}
d{\omega}\omega|\lambda^{0}_{\omega}V_{M,\omega}|^2\simeq
\frac{1}{2\pi}\int_{0}^{\infty}
d{\omega}\omega|g_{\omega}V_{M,\omega}|^2e^{-8\pi M\omega} \simeq
\frac{c}{128\pi^3 M^2}=cL_{l=0} \,,
\end{split}
\end{equation}
where $c=|g_{\omega}V_{M,\omega}|^2$ is assumed to be a model
dependent constant approximatively. $L_{l=0}$ is the $l=0$
luminosity (without backscattering effects), a little smaller than the result $(768\pi M^2)^{-1}$~\cite{j} based on the Hawking's arguments. The transition
rate for $\arrowvert0_M\rangle$ to all of the states
$\arrowvert\omega_0,\tilde{\omega}_0\rangle$ is given by
\begin{equation}
\label{eq:66}
\begin{split}
|\lambda^{1}_{\omega}V_{M,\omega}|^2\simeq
|g_{\omega}V_{M,\omega}|^2(\sum_{\omega_0}e^{-4\pi
M(\omega+\omega_0)})^2=|g_{\omega}V_{M,\omega}|^2\frac{e^{-8\pi
M\omega}}{8\pi^2 M^2} \,,
\end{split}
\end{equation}
reduced by a factor $(8\pi^2 M^2)^{-1}$ comparing with
$|\lambda^{0}_{\omega}V_{M,\omega}|^2$. This means higher states $\arrowvert n,\tilde{n}\rangle$ contribute little as long as the black hole mass is
large enough. However, it should be stressed again that the above calculations actually only provide a convenient computational method. Besides, the calculated transition rate in~\eqref{eq:65} serves only as the vacuum contribution indicated by the form of the state in~\eqref{eq:61}.

During the evolution of our model, the scalar field can be excited to higher levels, provided that the emitted energy of the black hole is still stored in the scalar field and not transferred into the detector. Then what are the effects of these excited states? This can be seen through the following two expectation values
\begin{subequations}\label{eq:66a}
\begin{align}
\label{eq:66a:1} \langle0_M\arrowvert N\arrowvert0_M\rangle =
\langle0_M\arrowvert\sum_{\omega}b_{\omega}^{\dag}b_{\omega}\arrowvert0_M\rangle=\sum_{\omega}\frac{1}{e^{8\pi
M\omega}-1} \,,
\\
\label{eq:66a:2} N(\omega_0)^{-1}\langle0_M\arrowvert
b_{\omega_0}Nb_{\omega_0}^{\dag}\arrowvert0_M\rangle=\sum_{\omega}\frac{1}{e^{8\pi
M\omega}-1}+2+\frac{e^{-8\pi M\omega_0}}{e^{8\pi M\omega_0}-1} \,,
\end{align}
\end{subequations}
with $b_{\omega_0}^{\dag}\arrowvert0_M\rangle$ a first excited state and $N(\omega_0)=e^{8\pi M\omega_0}(e^{8\pi M\omega_0}-1)^{-1}$
its normalization constant. We can see that, in addition to the thermal spectrum giving the vacuum effects, there is also a contribution $2+e^{-8\pi M\omega_0}(e^{8\pi M\omega_0}-1)^{-1}$. For a general excited state $(b)^m(b^{\dag})^n(\tilde{b})^p(\tilde{b}^{\dag})^q\arrowvert0_M\rangle$, the contributions will be more complex, but \emph{the thermal spectrum is modified to be non-thermal}. This implies that there are quantum corrections to the thermal spectrum in our model, which can be treated as an extension of the semiclassical method used by Hawking. In reference~\cite{k1}, the authors say that the non-thermal spectrum of Parikh and Wilczek~\cite{k} with the form~\eqref{eq:20} allows for the Hawking radiation emissions to carry away all information of a black hole\footnote{Notice that the state $\arrowvert0_{M-\omega}\rangle$ used in deriving~\eqref{eq:20} is also an excited state relative to $\arrowvert0_M\rangle$, as indicated by~\eqref{eq:59}. }. The calculations in~\eqref{eq:66a} support their argument only partly, because in this way not all information can be recovered within a black hole background. That is, the correlation information is still lost if local measurements are performed during the evaporation, as discussed below~\eqref{eq:f} or below~\eqref{eq:57}\footnote{From~\eqref{eq:21} and~\eqref{eq:22}, we can see that the discussion here is analogous to that below~\eqref{eq:57}. In particular, after the black hole evaporates completely, the $b_{\omega}$ modes will become $a_{\omega}$ modes, without involving any mode splitting. }. Below~\eqref{eq:57} a condition is proposed that measurements should be performed after the black hole evaporate completely. At that time, there will not be mode splitting for the scalar field, and ordinary measurement can be performed without any question.

Certainly, when the energy almost goes to the infinite future and is absorbed by
the detector, the scalar field can be in some near horizon vacuum state. Although the scalar field's (near horizon) vacuum is not
necessary during the evolution, it indeed
\emph{provides the necessary entanglements to generate the new correlations
between the black hole and the detector}. Then can we destroy the entanglements implicit in the vacuum $\arrowvert0_M\rangle$ by a Hawking's
measurement~\eqref{eq:39}? If this was possible, then
$\arrowvert0_M\rangle$ would collapse into a state like $\arrowvert\alpha\rangle\arrowvert\tilde{\beta}\rangle$, which behaves like two
independent systems described alone by $b_{\omega}$
and $\tilde{b}_{\omega}$ modes respectively. As a consequence, even the whole
space-time may be broken completely into two disconnected regions along the
event horizon. In our model, however, the local measurement is performed on the space of the detector,
then after the measurement it is the detector's state that will
collapse, so will the state of the black hole due to the entanglements between the black hole and the detector. These have little influences on the vacuum
$\arrowvert0_M\rangle$ without destroying the entanglements implicit in it, so the evolution or evaporation can be continuing.

\section{A Qualitative Analysis Including The Gravitational Perturbation }
\label{sec:nonlocal}

\subsection{Graviton as An Intermediate Medium For Energy Transfer}
\label{sec:gra}

The model studied in section~\ref{sec:model} provides a \emph{nonlocal correlation} between the causally disconnected interior and exterior of the black
hole. According to local QFT, the interactions $H_{B,\psi}$ and $H_{\psi,D}$ in~\eqref{eq:7} and~\eqref{eq:13} should decouple from each other. However, because of the entanglements between the $b_{\omega}$ and $\tilde{b}_{\omega}$ modes implicit in the vacuum state $\arrowvert0_M\rangle$, it is possible to combine the two interaction terms in a way like~\eqref{eq:14} to obtain a non-vanishing transition~\eqref{eq:17}, leading to correlations between those two disconnected regions. In this subsection, we shall make some further investigations in the framework of curved space QFT, where unlike the traditional semiclassical treatment, the gravity is treated by including
the gravitons or the perturbations of the black hole background.

For a flat space-time background, a useful expansion is given by
$g_{\mu\nu}\simeq \eta_{\mu\nu}+h_{\mu\nu}$, with the perturbation
$h_{\mu\nu}$ treated as the quantum gravitational field or graviton. Here, we
make an analogous expansion based on some black hole background
\begin{equation}
\label{eq:67}
\begin{split}
g_{\mu\nu}\simeq g^{B}_{\mu\nu}+h_{\mu\nu} \,,
\end{split}
\end{equation}
with $g^{B}_{\mu\nu}$ the classical metric of a black hole with mass $M_0$,
and the perturbation $h_{\mu\nu}$ treated still as the quantum gravitational field based on this black hole. Analogous
to the mode expansion~\eqref{eq:9} for the scalar field, we
have\footnote{For simplicity, we still use the formulae of the shock wave model, so the coordinate transformations for the tensor can be ignored.}
\begin{equation}
\label{eq:68}
\begin{split}
h_{\mu\nu}=\int_{0}^{\infty}
d\omega\sum_i\epsilon^{i}_{\mu\nu}(\omega)(e^{i}_{\omega}U_{\omega}+h.c.)=\int_{0}^{\infty}
d\omega\sum_i\epsilon^{i}_{\mu\nu}(\omega)[c^{i}_{\omega}u_{\omega}+\tilde{c}^{i}_{\omega}\tilde{u}_{\omega}+h.c.]
\,,
\end{split}
\end{equation}
with $\epsilon_{\mu\nu}$ the polarizations of the gravitons. The full interaction is given by
\begin{equation}
\label{eq:69}
\begin{split}
\int_{\Sigma_{int}}
d^3xh^{\tilde{c}}_{\mu\nu}T^{_{\mu\nu}}[\psi_{\tilde{b}}]+\int_{\Sigma_{ext}}
d^3xh^{c}_{\mu\nu}T^{_{\mu\nu}}[\psi_b] \,,
\end{split}
\end{equation}
which occurs in both of the interior and exterior of the black hole according to the extended BHC (ii'). For simplicity, the interactions will be studied only in the momentum
space labeled by the energy $\omega$, with all the other items such
as the polarizations being neglected. According to local QFT, the terms in~\eqref{eq:69} will
contain the following interaction patterns
\begin{equation}
\label{eq:70}
\begin{split}
(b_{\omega_2}^{\dag}c_{\omega}b_{\omega_1}+h.c.)\delta(\omega_1+\omega-\omega_2)\qquad
(c_{\omega}^{\dag}b_{\omega_1}b_{\omega_2}+h.c.)\delta(\omega_1+\omega_2-\omega)\\
(\tilde{b}_{\omega_2}^{\dag}\tilde{c}_{\omega}\tilde{b}_{\omega_1}+h.c.)\delta(\omega_1+\omega-\omega_2)\qquad
(\tilde{c}_{\omega}^{\dag}\tilde{b}_{\omega_1}\tilde{b}_{\omega_2}+h.c.)\delta(\omega_1+\omega_2-\omega)
\,,
\end{split}
\end{equation}
with the combinations of the mode functions such as $\int u^*\partial u\partial u$ being ignored for simplicity.

Let's then evolve the entire system, and a Hartree-Fock-like method should be used. For an initial state $\arrowvert i\rangle_g$ of the graviton, we can
evolve it and obtain a final state $\arrowvert f\rangle_g$ via the evolution operator for the interaction in~\eqref{eq:69}, where the fields are expanded
based on an initial background metric $g^{B}_{\mu\nu}$ as in~\eqref{eq:9} and~\eqref{eq:68}. Then the new background metric is
given by $g^{B}_{\mu\nu}+_g\langle f\arrowvert h_{\mu\nu}\arrowvert
f\rangle_g$, and the procedure continues. Let's first consider the black hole formation. The initial state $\arrowvert i\rangle_g$ of the graviton
must be chosen to satisfy $_g\langle i\arrowvert
h_{\mu\nu}\arrowvert i\rangle_g\simeq0$, so that the black hole is initially stationary.
To form a larger black hole with mass $M_0+\Delta M=M$, the scalar field must be in its excited state initially, for example\footnote{Here $\prod_{\omega}a_{\omega}^{\dag}\equiv
a_{\omega_1}^{\dag}a_{\omega_2}^{\dag}\cdots$, $\sum\omega\equiv\omega_1+\omega_2+\cdots$, similarly for other equations below, such as~\eqref{eq:72},~\eqref{eq:74} and~\eqref{eq:76}, etc.}
\begin{equation}
\label{eq:71}
\begin{split}
\arrowvert\phi\rangle_{\psi}\sim
\prod_{\omega}a_{\omega}^{\dag}\arrowvert0_{M_0}\rangle_{\psi}\qquad
(\sum\omega=\Delta M) \,.
\end{split}
\end{equation}
Because of the Bogoliubov transformations, this state can be
expressed as a superposition of various states, among which two
particular ones are
\begin{equation}
\label{eq:72}
\begin{split}
\arrowvert\phi_1\rangle_{\psi}\sim
\prod_{\omega_1}b_{\omega_1}^{\dag}\arrowvert0_{M_0}\rangle_{\psi}(\sum\omega_1=\Delta
M_1),\qquad \arrowvert\phi_2\rangle_{\psi}\sim
\prod_{\omega_2}\tilde{b}_{\omega_2}^{\dag}\arrowvert0_{M_0}\rangle_{\psi}(\sum\omega_2=\Delta
M_2) \,.
\end{split}
\end{equation}
After the formation of the larger black hole, the state of the
scalar field can be chosen to be the vacuum
$\arrowvert0_M\rangle_{\psi}$. That is, the energy $\Delta M$
stored in $_{\psi}\langle\phi\arrowvert
T_{\mu\nu}[\psi_{b,\tilde{b}}]\arrowvert\phi\rangle_{\psi}$ have been almost
transferred into the black hole (interior), keeping the energy
balance\footnote{Here the first order Einstein's equation has been used, since $R_{\mu\nu}[g^{B}]=0$. The energy balance is a general condition satisfied by two components of a closed system.}
\begin{equation}
\label{eq:73}
\begin{split}
_g\langle f\arrowvert G_{\mu\nu}[h_{c,\tilde{c}}]\arrowvert
f\rangle_g\sim  _{\psi}\langle\phi\arrowvert
T_{\mu\nu}[\psi_{b,\tilde{b}}]\arrowvert\phi\rangle_{\psi} \,.
\end{split}
\end{equation}
The graviton's final state can be chosen as, for example (according to BHC
(iii))
\begin{equation}
\label{eq:74}
\begin{split}
\arrowvert
f\rangle_g\sim\prod_{\omega}\tilde{c}_{\omega}^{\dag}\arrowvert
i\rangle_g\qquad (\sum_{\omega}\omega=\Delta M) \,,
\end{split}
\end{equation}
since the energy of the black hole must be stored almost in the interior degrees of freedom. Then there would be some transitions due to the interactions in
\eqref{eq:69}
\begin{equation}
\label{eq:75}
\begin{split}
\arrowvert\phi_1\rangle_{\psi}\arrowvert
i\rangle_g\stackrel{1}{\rightarrow}
\arrowvert0_M\rangle_{\psi}\arrowvert f_1\rangle_g,\qquad
\arrowvert\phi_2\rangle_{\psi}\arrowvert
i\rangle_g\stackrel{2}{\rightarrow}
\arrowvert0_M\rangle_{\psi}\arrowvert f_2\rangle_g \,,
\end{split}
\end{equation}
with energies $\Delta M_1$ and $\Delta M_2$ for $\arrowvert
f_1\rangle_g$ and $\arrowvert f_2\rangle_g$, respectively.

The transition 2 in~\eqref{eq:75} can simply be induced by an interaction
$\tilde{c}_{\omega}^{\dag}\tilde{b}_{\omega_1}\tilde{b}_{\omega_2}$
in~\eqref{eq:70}, which is local in the interior of the black hole. While for the transition 1, it
seems that an interaction $\tilde{c}_{\omega}^{\dag}b_{\omega_1}b_{\omega_2}$ can do the job, which is
impossible since it contains operators from causally disconnected regions. But following the effective field model in section~\ref{sec:model}, the transition 1 can be induced by some combinations of some local interaction terms\footnote{We have neglected the mass
dependence of these operators for simplicity.}
\begin{subequations}\label{eq:76}
\begin{align}
\label{eq:76:1} \prod_{\sum\omega'=\Delta
M_1}(\tilde{c}_{\omega'}^{\dag}\tilde{b}_{\omega'_1}\tilde{b}_{\omega'_2})(c_{\omega'}b_{\omega'_1}b_{\omega'_2})\prod_{\sum\omega=\Delta
M_1}(c_{\omega}^{\dag}b_{\omega_1}b_{\omega_2})\,,
\\
\label{eq:76:2} \prod_{\sum\omega'=\Delta
M_1}(\tilde{c}_{\omega'}^{\dag}\tilde{b}_{\omega'_1}^{\dag}\tilde{b}_{\omega'_2}^{\dag})(c_{\omega'}b_{\omega'_1}^{\dag}b_{\omega'_2}^{\dag})\prod_{\sum\omega=\Delta
M_1}(c_{\omega}^{\dag}b_{\omega_1}b_{\omega_2})\,.
\end{align}
\end{subequations}
That is, the energy of the scalar field is first transferred
into the exterior gravitational field via some local
interactions, then into the interior gravitational field through the
entanglements implicit in $\arrowvert0_{M_0}\rangle_{\psi}$. These can be
verified by noting the following actions
\begin{subequations}\label{eq:77}
\begin{align}
\label{eq:77:1}
(\tilde{b}_{\omega'}b_{\omega'})(b_{\omega}b_{\omega}^{\dag})\arrowvert0_{M_0}\rangle_{\psi}\sim(\tilde{b}_{\omega'}b_{\omega'})\arrowvert0_{M_0}\rangle_{\psi}\sim\arrowvert0_M\rangle_{\psi}\,,
\\
\label{eq:77:2}
(\tilde{b}_{\omega'}^{\dag}b_{\omega'}^{\dag})(b_{\omega}b_{\omega}^{\dag})\arrowvert0_{M_0}\rangle_{\psi}\sim(\tilde{b}_{\omega'}^{\dag}b_{\omega'}^{\dag})\arrowvert0_{M_0}\rangle_{\psi}
\sim\arrowvert0_M\rangle_{\psi}\,,
\\
\label{eq:77:3}
\tilde{c}_{\omega'}^{\dag}(c_{\omega}c_{\omega}^{\dag})\arrowvert
i\rangle_g\sim\tilde{c}_{\omega'}^{\dag}\arrowvert i\rangle_g\,,
\end{align}
\end{subequations}
where the creators and annihilators of the same mode has been considered to be approximately cancelled\footnote{Since $b,\tilde{b}$ can not annihilate $\arrowvert0_{M_0}\rangle_{\psi}$, thus $b_{\omega}b_{\omega}^{\dag}\arrowvert0_{M_0}\rangle_{\psi}=\arrowvert0_{M_0}\rangle_{\psi}+b_{\omega}^{\dag}b_{\omega}\arrowvert0_{M_0}\rangle_{\psi}$. The first term $\arrowvert0_{M_0}\rangle_{\psi}$ is already in~\eqref{eq:77:1} and~\eqref{eq:77:2}, while the second term can be considered to be from another procedure involving a state $\arrowvert\phi'_1\rangle_{\psi}\sim\prod_{\omega_1}b_{\omega_1}\arrowvert0_{M_0}\rangle_{\psi}$, which can also be derived from the general state in~\eqref{eq:71}.}, while those of different modes are in pairs so that the entanglements between the $b_{\omega}$ and $\tilde{b}_{\omega}$ modes are still retained. There are two interaction terms which do not appear in the ordinary patterns
\eqref{eq:70}
\begin{equation}
\label{eq:78}
\begin{split}
\tilde{c}_{\omega}^{\dag}\tilde{b}_{\omega_1}^{\dag}\tilde{b}_{\omega_2}^{\dag},\qquad
c_{\omega}b_{\omega_1}b_{\omega_2} \,.
\end{split}
\end{equation}
In the framework of flat space QFT, these two terms are related to a factor $\delta(\omega_1+\omega_2+\omega)$, and thus should vanish. In the black hole physics, however, the whole space-time is divided into an interior and an exterior regions by the event horizon, with each one \emph{incomplete}. As a consequence, the integral intervals of the space-time integrals in each region are also incomplete, leading
to non-vanishing results. Therefore, with the help of the combinations in~\eqref{eq:76}, the transition 1 in~\eqref{eq:75} can be induced step by step.
The other states in the superposition of $\arrowvert\phi\rangle_{\psi}$
can be treated in an analogous way.

These transitions can be illustrated in the following diagram

\[
\begin{CD}
b @>(\times )>> \tilde{b} \\
@V{bbc^{\dag}+h.c.}VV @VV{\tilde{b}\tilde{b}\tilde{c}^{\dag}+h.c.}V \\
c @>{\rm E(b,\tilde{b})}>{\rm c^{\dag}\tilde{c}+h.c.}> \tilde{c}
\end{CD}
\]
with $E(b,\tilde{b})$ denoted as the entanglements
implicit in the vacuum $\arrowvert0_{M_0}\rangle_{\psi}$. The transferred energy flows
along the arrows. First, it is stored in the scalar field as
matter's energy. For the energy stored in the $\tilde{b}_{\omega}$ modes,
it can be transferred directly into the $\tilde{c}_{\omega}$ modes as the
black hole's mass via the local interaction
$(\tilde{b}\tilde{b}\tilde{c}^{\dag}+h.c.)$. While for the energy
stored in the $b_{\omega}$ modes, it firstly has to be transferred
into the $c_{\omega}$ modes via the local interaction
$(bbc^{\dag}+h.c.)$, then into the $\tilde{c}_{\omega}$ modes as the
black hole's mass via the nonlocal correlation
$(c^{\dag}\tilde{c}+h.c.)$ that is generated by
$E(b,\tilde{b})$. This generated nonlocal correlation is
just like the one discussed in section~\ref{sec:infor}, by
noting $\tilde{c}_{\omega'}^{\dag}c_{\omega'}(_{\psi}\langle0_{M_0}\arrowvert\tilde{b}_{\omega'_1}\tilde{b}_{\omega'_2}b_{\omega'_1}b_{\omega'_2}\arrowvert0_{M_0}\rangle_{\psi})$ from~\eqref{eq:76}. The notation $(\times)$ in the above diagram is used to indicate the impossibility to transfer energy in the corresponding direction\footnote{In fact, there is an analogous way to transfer energy between the $\tilde{b}_{\omega}$ and $b_{\omega}$ modes via the entanglements between the $\tilde{c}_{\omega}$ and $c_{\omega}$
modes of the graviton's states. However, this seems to be impossible here since we have to make an average $\langle h\rangle$ to give a classical background in each stage.}, similarly for another diagram below.

For the black hole evaporation, we can reverse the above
formation process, with the initial and final states given by
$\arrowvert0_M\rangle_{\psi}\arrowvert f\rangle_g$ and $\arrowvert\phi\rangle_{\psi}\arrowvert i\rangle_g$, respectively\footnote{This procedure for the evaporation, by directly reversing the previous formation process, is a little like the stimulated radiation, because the state of the gravitons has been excited after the previous formation. }. The required
transitions can be induced by means of a local interaction
$\tilde{c}_{\omega}\tilde{b}_{\omega_1}^{\dag}\tilde{b}_{\omega_2}^{\dag}$
and the following combinations of local interactions
\begin{subequations}\label{eq:79}
\begin{align}
\label{eq:79:1} \prod_{\sum\omega'=\Delta
M_1}(c_{\omega'}b_{\omega'_1}^{\dag}b_{\omega'_2}^{\dag})\prod_{\sum\omega=\Delta
M_1}(c_{\omega}^{\dag}b_{\omega_1}^{\dag}b_{\omega_2}^{\dag})(\tilde{c}_{\omega}\tilde{b}_{\omega_1}^{\dag}\tilde{b}_{\omega_2}^{\dag})\,,
\\
\label{eq:79:2} \prod_{\sum\omega'=\Delta
M_1}(c_{\omega'}b_{\omega'_1}^{\dag}b_{\omega'_2}^{\dag})\prod_{\sum\omega=\Delta
M_1}(c_{\omega}^{\dag}b_{\omega_1}b_{\omega_2})(\tilde{c}_{\omega}\tilde{b}_{\omega_1}\tilde{b}_{\omega_2})\,.
\end{align}
\end{subequations}
Then the energy stored in
the interior of the black hole can be transferred back into the scalar field, and the scalar field will be excited. In
order for the black hole to evaporate further, the energy stored in
the scalar field must be transferred into somewhere else so that the
entanglements $E(b,\tilde{b})$ can still be utilized. The distance
detector in the model of section~\ref{sec:model} just does this job. These can be
illustrated in an analogous diagram

\[
\begin{CD}
 \tilde{b} @>(\times)>> b@>{\rm b^{\dag}d+h.c.}>> d\\
 @V{\tilde{b}\tilde{b}\tilde{c}^{\dag}+h.c.}VV @AA{bbc^{\dag}+h.c.}A \\
 \tilde{c} @>{\rm E(b,\tilde{b})}>{\rm c^{\dag}\tilde{c}+h.c.}> c
\end{CD}
\]
where we have reversed the direction for the energy flowing between the
$\tilde{b}_{\omega}$ and $\tilde{c}_{\omega}$ modes so that the energy can almost be transferred into the distant detector.
The role played by the gravitons as an intermediate medium is thus
demonstrated in the above diagram, or more apparently in the following expression
\begin{equation}
\label{eq:80}
\begin{split}
\prod_{\sum(\omega_1+\omega_2)=\Delta
M_1}(d_{\omega_1}^{\dag}d_{\omega_2}^{\dag}b_{\omega_1}b_{\omega_2})\Big\{\prod_{\sum\omega'=\Delta
M_1}(c_{\omega'}b_{\omega'_1}^{\dag}b_{\omega'_2}^{\dag})\prod_{\sum\omega=\Delta
M_1}(c_{\omega}^{\dag}b_{\omega_1}b_{\omega_2})\Big\}_{g,\psi}(\tilde{c}_{\omega}\tilde{b}_{\omega_1}\tilde{b}_{\omega_2})
\,.
\end{split}
\end{equation}
The grouped contribution from the scalar and gravitational field
serves as the propagators in the exterior region, while the terms in
the two ends just give the model in section~\ref{sec:model} with an
identification $\tilde{c}_{\omega}\rightarrow \mathbf{V}_{\omega}$.
Roughly speaking, the energy (or information) is not transferred instantaneously from the interior of the black hole into the distant detector, but transferred step by step via the intermediate gravitational and scalar field. Therefore, the correlations between the black hole and the detector are physically practicable.

As discussed in section~\ref{sec:bhc}, in the (classical) black hole physics, there is a contradiction between the descriptions of a static observer and an in-falling observer because of the metric singularity at $r=2M$. In a quantum version, as illustrated in
the above two diagrams, it is impossible for an exterior or interior particle to cross
the event horizon directly, since the exterior particle is described by the $b_{\omega}$ modes, while the interior one is described in terms of the
$\tilde{b}_{\omega}$ modes. This causally disconnectedness agrees with the classical one in the view of a static observer, since the chosen
reference frame is still given by $(t,r,\theta,\phi)$. For an in-falling observer, the chosen reference frame is regular so that
there is no mode split. Then the physical world in his
eyes can be described well according to flat space QFT.
The scalar field and the gravitational perturbation are expanded in terms of the $a_{\omega}$ and $e_{\omega}$ modes respectively, in particular, the vacuum for the scalar field is
$\arrowvert0_M\rangle$. The full interaction is given by the term~\eqref{eq:27}, which can induce various quantum transitions, including those induced by nonlocal correlations when expressed in terms of the $b_{\omega}$ and $\tilde{b}_{\omega}$ modes. For example, by means of the Bogoliubov transformations, an
interaction pattern $e_{\omega}^{\dag}a_{\omega_1}a_{\omega_2}$ can be formally expressed as
\begin{equation}
\label{eq:81}
\begin{split}
e_{\omega}^{\dag}a_{\omega_1}a_{\omega_2}\sim \alpha
c_{\omega'}^{\dag}b_{\omega'_1}b_{\omega'_2}+
\beta\tilde{c}_{\omega'}^{\dag}b_{\omega'_1}b_{\omega'_2}+\cdots \,,
\end{split}
\end{equation}
where the term $c_{\omega'}^{\dag}b_{\omega'_1}b_{\omega'_2}$ is an local interaction in the exterior region, while the
other one $\tilde{c}_{\omega'}^{\dag}b_{\omega'_1}b_{\omega'_2}$ is a nonlocal one which is required to induce the
transition 1 in~\eqref{eq:75}.

For a static observer, however, the term $\tilde{c}_{\omega'}^{\dag}b_{\omega'_1}b_{\omega'_2}$ in~\eqref{eq:81} can not be constructed directly since it contains modes from
disconnected regions, violating the causality. But the quantum
transitions are physical so that they should not depend on the observers. These quantum transitions, in the view of a static observer, are given by~\eqref{eq:76} and~\eqref{eq:79} via the entanglements implicit in the vacuum $\arrowvert0_M\rangle$. That is, they are induced by some nonlocal correlations between the interior and exterior degrees of freedom of the black hole, leading to \emph{some quantum tunneling effects across the event horizon}. Thus, the energy or information can be tunneled across the event horizon gradually through those quantum transitions, as long as the entanglements implicit in $\arrowvert0_M\rangle_{\psi}$ will never be destroyed.
The remaining problem is whether the different descriptions
are consistent with each other, one is based on the static observer, the other is based on the in-falling observer. For instance, on one hand,
$\tilde{c}_{\omega}^{\dag}b_{\omega_1}b_{\omega_2}$, as one part of the
full interaction in~\eqref{eq:81}, can not be constructed directly for a static observer; on the other hand, terms like
$e_{\omega}a_{\omega_1}a_{\omega_2}$ should be vanishing for an
in-falling observer, while terms like
$c_{\omega}b_{\omega_1}b_{\omega_2}$ or
$\tilde{c}_{\omega}\tilde{b}_{\omega_1}\tilde{b}_{\omega_2}$ can be
constructed for a static observer. This consistency problem has been answered in section~\ref{sec:bhc} only in principle by means of the principle of general covariance, without a detailed inspection\footnote{This consistency problem can be treated as an extension of the correspondence~\eqref{eq:c}. However, as shown in section~\ref{sec:bhc}, general covariance is violated in the framework of effective field theory, thus the consistency problem may be resolved only in a quantum gravity theory.}.

\subsection{Three Classes of Entanglements or Correlations}
\label{sec:entan}

Up to now, we have obtained three kinds of entanglements:

(1) Entanglements that are always implicit in some steady pure state, such as the
vacuum state $\arrowvert0_{M}\rangle$;

(2) Entanglements that can be established through some ordinary local
interactions, like those $(bbc^{\dag}+h.c.)$,
$(\tilde{b}\tilde{b}\tilde{c}^{\dag}+h.c.)$ and $(bd^{\dag}+h.c.)$;

(3) Entanglements which are \emph{nonlocal} in the sense that they are established between two causally disjoint regions by means of the entanglement implicit in the
vacuum state $\arrowvert0_{M}\rangle$, like
$(c^{\dag}\tilde{c}+h.c.)$ and $(d^{\dag}\mathbf{V}+h.c.)$.

From the previous discussions, one can see that the entanglements of class (1) and (3) share some property different from that of the class (2): \emph{local measurements will lead to (correlation) information loss inevitably due to the causal disconnectedness}. While for the entanglements in the class (2), correlation information can be acquired by comparing two local measurement outcomes through a classical channel, such as the correlation $\langle S_1S_2\rangle$ for an EPR pairs. The author of reference~\cite{g} tried to transfer the entanglements of class (1) into those among the radiations that belong to the class (2). In our model, the information is stored in the entanglements of class (3) between the causally disconnected interior and exterior of the black hole. According to the above properties of the three classes, it seems that both the two mechanisms are impossible. Then how to make a distinction between the generation of entanglements in our model and the transfer of entanglements in reference~\cite{g}. There is still a difference between the classes (1) and (3) by noting that $\arrowvert0_{M}\rangle$ is a steady pure state, while the entanglements of class (3) are only \emph{established temporarily}. This can be explained as follows. In our model the energy of the radiation mainly comes from the black hole, transferred via the the entanglements of class (3). This means that the entanglements of class (3) will disappear when all the energy of the black hole is transferred outside completely, so that the black hole is also absent. At that time, the class (3) disappear while vacuum state still remains, and entanglements will always be established via local interactions, belonging to class (2). In this sense, our entanglement generation mechanism is more realizable than the transfer of entanglements in reference~\cite{g}.

Let's now consider the changes of the vacuum
$\arrowvert0_{M}\rangle$ during the evolution.
In section~\ref{sec:model}, a restrict space $\mathcal {H}_0$ is assigned, in which the scalar
field are assumed to be in the (near horizon) vacua depending on different black hole's masses. As
the entire system evolves, a sequence of those vacua~\eqref{eq:33:3} is assigned to describe the changes of the vacuum during the evolution. It seems to be impossible to decide which vacuum has more entanglements, since each vacuum can be expressed by~\eqref{eq:4} in terms of their
respective $b_{\omega}, \tilde{b}_{\omega}$ modes, i.e. maximum entangled in each Fock space. If
$\arrowvert0_{M_0}\rangle$ is a maximum entangled state, then
$\arrowvert0_{M_1}\rangle$ is away from maximum by including excited
states from the relation~\eqref{eq:59}. But this relation can also be rewritten in a reversed
order so that $\arrowvert0_{M_1}\rangle$ is maximum entangled while
$\arrowvert0_{M_0}\rangle$ is away from maximum. One may say that their entanglements can be compared by calculating the Von Neumann entropy for each vacuum. But this is complicated due to the mass dependence in each vacuum $\arrowvert0_{M}\rangle$. In fact, those vacua are just meta-stable states, and should be treated as some auxiliary states. For a full evolution, free states in the real $t\rightarrow\pm\infty$ regions should be utilized, since the causal structures of the space-time there should be regular without black holes and mode splitting. In this sense, it's meaningless to count the amount of entanglements implicit in those vacua.

For local interactions in terms of perturbation expansions, we can obtain states with finite creators and annihilators acting on the vacuum, which are still entangled states, for example a state like $(b)^m(b^{\dag})^n(\tilde{b})^p(\tilde{b}^{\dag})^q\arrowvert0_M\rangle$. That is, the entanglements between the $b_{\omega}$ and $\tilde{b}_{\omega}$ modes are still remaining, since $\arrowvert0_{M}\rangle$ can not be destroyed by those operations composing of only finite creators and annihilators. This is not the case for a global operation, for example the inverse of the operator $e^{\sum_{\omega}e^{-4\pi M\omega}b_{\omega}^{\dag}\tilde{b}_{\omega}^{\dag}}$ in~\eqref{eq:4} that can be used to transform $\arrowvert0_{M}\rangle$ to another vacuum $\arrowvert0,\tilde{0}\rangle$, obtaining a direct product state. Fortunately, this kind of operators with global properties seldom occurs in local effective field theory. This can also be explained by noting that the local interactions $H_{B,\psi}$ and $H_{\psi,D}$ are independent so that there exist no prior correlations for them to produce an operator with global properties. Therefore, without destroying the entanglements implicit in $\arrowvert0_{M}\rangle$, the quantum transitions given in sections~\ref{sec:model} and~\ref{sec:gra} can still continue, leading to the black hole
evaporation. It thus implies that the space $\mathcal {H}_0$ should be extended to a more general one to \emph{include all the possible entangled states of the scalar field, generated by those operations composing of finite creators and annihilators acting on each vacuum state $\arrowvert0_{M}\rangle$}, analogous to the case of the four Bell states. This can be explicitly seen in the qubit model analyzed in appendix~\ref{sec:qubit}, where all the four Bell states can appear during the unitary evolution.

\section{Summary}
\label{sec:sum}

Some essential points for our model of the black hole evaporation are summarized as follows:

(1) Black hole complementarity (BHC) can almost be satisfied, but with an extended postulate (ii'): both the exterior and interior regions of the event horizon are well described by QFT in curved space, with the singularity $r=0$ excluded from the interior region. (Sections~\ref{sec:Model},~\ref{sec:dynamics} and~\ref{sec:nonlocal}.) The reason for the extension of the BHC (ii) is given in section~\ref{sec:bhc}.

(2) The radiation mainly comes from the black hole itself, transferred outside via some quantum
transitions across the event horizon. Those quantum transitions are induced via some nonlocal correlations
generated by the entanglements implicit in the (near horizon) vacuum state $\arrowvert0_{M}\rangle$ of the
scalar filed. The contribution from the stretched horizon or the near horizon region can be treated as vacuum effects.
 (Sections~\ref{sec:model},~\ref{sec:infor},~\ref{sec:gra} and~\ref{sec:ext})

(3) A Hawking-like measurement $\langle0_M\arrowvert\mathcal{O}_{ext}\arrowvert0_M\rangle$ will always lead to correlation information loss. In our model, Hawking-like measurement can be replaced by measurements operated on an added detector that receives radiations. In this way, instead of Hawking's
thermal spectrum, a non-thermal spectrum of the radiations will be obtained, by including the contributions from the possible intermediate states. (Sections~\ref{sec:bhc},~\ref{sec:interaction},~\ref{sec:infor} and~\ref{sec:mix})

(4) Information will be lost when some local measurements
have been performed. This is caused by quantum decoherence
that destroys the entanglements among the components of a
closed system. Moreover, in the framework of effective field theory, measurements should be performed only after the black hole evaporates completely, otherwise correlation information will be lost for ever. A quantum gravity theory without background dependence may resolve this problem. (Sections~\ref{sec:mix},~\ref{sec:bhc})

(5) In the framework of curved space QFT, a unitary evolution including both the black hole formation and evaporation can be constructed formally, although the constructed evolution operators have singularities. Those singularities can not be easily avoided, indicating that a quantum gravity theory is still needed. (Section~\ref{sec:S} and Appendix~\ref{sec:singular})

Here, we give a qubit model which behaves in a similar manner as the above
essential points. Suppose that Alice and Bob own an EPR pairs, Alice
takes one qubit denoted by $a$, while Bob carries the other one named by
$b$. Now there is a task for them to establish correlations between two independent systems $A$ and $B$,
which are far away from each other. Usually, for two systems to
correlate with each other, one direct method is to couple them via
some local interactions, for example an EPR pairs can be produced in this way.
However, if the systems $A$ and $B$ can not be moved to close to
each other, or even they may be located in two causally disconnected regions, for instance the interior and exterior of a black hole, then how to correlate them? The EPR pairs owned by Alice and Bob can be applied to accomplish this task in the following way.
Let Alice and Bob travel to the locations of $A$ and $B$ respectively,
together with their own qubits of the EPR pairs. After arriving, let them carry some local unitary
operations $U_{aA}$ and $U_{bB}$ on the combined systems $aA$ and $bB$ respectively, then $A$ and $B$ will be correlated with each other by carefully controlling the operations.
The mathematical detail of this model is given in appendix~\ref{sec:qubit}, where a \emph{modified quantum teleportation} is also given to transfer information through the EPR pairs.
Obviously, the systems $A$ and $B$ can be regarded as the black hole and the detector respectively, while the EPR pairs serves as the vacuum state $\arrowvert0_{M}\rangle$. Therefore, the information of the black hole can also be transferred outside via an analogous quantum teleportation process.

From the discussions of this paper, it can be concluded that, the black hole formation and evaporation processes can be modelled in a unitary manner according to effective field theory. Then the paradoxes of the information loss and the firewall can be resolved, provided that no local measurements are performed during the process. Even
though the black hole may evaporate completely in the end, the
information could still be stored in the entire system, with components including the scalar field, the weak gravitational field (without black holes), and perhaps an additional detector. When some local measurement has been performed, however, the information will be lost inevitably due to quantum decoherence, but lost only partially since contributions from the intermediate states have been included. A more general case including the charges and angular momenta has to be further investigated. Since our model is unitary, and a non-thermal spectrum is also obtained indicated by~\eqref{eq:57}, it thus seems that \emph{there may not be a thermodynamic character for the pure gravity}. Certainly, this needs to be investigated further. In conclusion, although our model is just an approximation, it indeed implies that quantum mechanics and gravity can be combined in a consistent way, giving a quantum gravity theory.

\appendix

\section{ A Simple Model With a Singular Evolution Operator}
\label{sec:singular}

A simple model with a singular evolution operator is studied in detail here. The action is that for a
harmonic oscillator with a prescribed, time-dependent spring
``constant''~\cite{l}
\begin{equation}
\label{eq:82}
\begin{split}
S[q] &=
\int[-\frac{1}{2}\dot{q}^2(t)+\frac{1}{2}\omega^2(t)q^2(t)]dt \,,
\end{split}
\end{equation}
where
\begin{equation}
\label{eq:83}
\begin{split}
\omega^2(t) &= A+B\tanh\lambda t \,.
\end{split}
\end{equation}
The function $\omega^2(t)$ becomes constant in the remote past and
the remote future
\begin{equation}
\label{eq:84}
\begin{split}
\omega^2(t)\stackrel{t\rightarrow\pm \infty}{\longrightarrow}A\pm
B\equiv \omega_{\pm }^2 \,.
\end{split}
\end{equation}
By solving the equations of motion, the in and out
mode functions can be obtained~\cite{l}. As in curved space QFT,
these two modes are related by a Bogoliubov transformation.

From~\eqref{eq:84},the free Hamiltonians in the $t\rightarrow\pm
\infty$ limits are
\begin{equation}
\label{eq:85}
\begin{split}
H\stackrel{t\rightarrow\pm \infty}{\longrightarrow} H^{\pm
}=\frac{1}{2}(p^2+\omega_{\pm }^2q^2) \,,
\end{split}
\end{equation}
with the eigenstates given by
\begin{equation}
\label{eq:86}
\begin{split}
H^{\pm }\arrowvert\chi^{\pm}_\alpha\rangle &= E^{\pm
}_\alpha\arrowvert\chi^{\pm }_\alpha\rangle \,.
\end{split}
\end{equation}
The ``out''($+$) and ``in''($-$) states are defined as
\begin{equation}
\label{eq:87}
\begin{split}
H\arrowvert\Phi^{\pm }_\alpha\rangle &= E^{\pm
}_\alpha\arrowvert\Phi^{\pm }_\alpha\rangle \,,
\end{split}
\end{equation}
which satisfy the condition~\cite{m}
\begin{equation}
\label{eq:88}
\begin{split}
\exp(-i H t)\sum_{\alpha}g(\alpha)\arrowvert\Phi_{\alpha}^{\pm
}\rangle\stackrel{t\rightarrow\pm \infty}{\longrightarrow} \exp(-i
H^{\pm }t)\sum_{\alpha}g(\alpha)\arrowvert\chi^{\pm }_\alpha\rangle
\,.
\end{split}
\end{equation}
Then the ``out'' and ``in'' states can be expressed as
\begin{equation}
\label{eq:89}
\begin{split}
\arrowvert\Phi_{\alpha}^{\pm }\rangle=\Omega^{\pm} (\pm
\infty)\arrowvert\chi^{\pm }_\alpha\rangle \,,
\end{split}
\end{equation}
where
\begin{equation}
\label{eq:90}
\begin{split}
\Omega^{\pm }(t)\equiv\exp(+i H t)\exp(-i H^{\pm }t) \,.
\end{split}
\end{equation}
The S-matrix is thus given by
\begin{equation}
\label{eq:91}
\begin{split}
(S_{+-})_{\beta\alpha} &=
\langle\Phi_{\beta}^{+}\arrowvert\Phi_{\alpha}^{-}\rangle \,,
\end{split}
\end{equation}
or equivalently the S-operator is
\begin{equation}
\label{eq:92}
\begin{split}
S_{+-} &= \Omega^+(+\infty)^{\dag}\Omega^-(-\infty) =
U_{+-}(+\infty,-\infty) \,,
\end{split}
\end{equation}
where the evolution operator is defined as
\begin{equation}
\label{eq:93}
\begin{split}
U_{+-}(t_2,t_1)\equiv \Omega^+(t_2)^{\dag}\Omega^-(t_1)=\exp(+i
H^+t_2)\exp\{-i H(t_2-t_1)\}\exp(-i H^-t_1) \,.
\end{split}
\end{equation}

Differentiating $U_{+-}(t_2,t_1)$ with respect to $t_2$ and $t_1$ respectively, we have
\begin{subequations}\label{eq:94}
\begin{align}
\label{eq:94:1}
i\frac{d}{dt_{2}}U_{+-}(t_2,t_1)=V^+(t_2)U_{+-}(t_2,t_1)\qquad
V^+(t_2)=e^{iH^+t_2}(H-H^+)e^{-iH^+t_2}\,,
\\
\label{eq:94:2}
-i\frac{d}{dt_{1}}U_{+-}(t_2,t_1)=U_{+-}(t_2,t_1)V^-(t_1)\qquad
V^-(t_1)=e^{iH^-t_1}(H-H^-)e^{-iH^-t_1}\,.
\end{align}
\end{subequations}
Since $H^+\neq H^-$, then $V^+(t)\neq  V^-(t)$ generally, meaning that the first differential of
$U_{+-}(t_2,t_1)$ is discontinuous. This can also bee seen from a singular initial condition
\begin{equation}
\label{eq:95}
\begin{split}
U_{+-}(t_0,t_0) &= e^{iH^+t_0}e^{-iH^-t_0} \,,
\end{split}
\end{equation}
indicating that $U_{+-}(t_2,t_1)$ works well only if $t_2\neq t_1$. $U_{+-}(t_2,t_1)$ occurs in the two-point
function $\langle0^+\arrowvert q(t)q(t')\arrowvert0^-\rangle$, while
for $\langle0^+\arrowvert q(t)q(t')\arrowvert0^+\rangle$ and
$\langle0^-\arrowvert q(t)q(t')\arrowvert0^-\rangle$, the evolution
operators $U_{++}$ and $U_{--}$ are
\begin{equation}
\label{eq:96}
\begin{split}
U_{\pm\pm}(t_2,t_1) &= \exp(+i H^{\pm} t_2)\exp\{-i
H(t_2-t_1)\}\exp(-i H^{\pm}t_1) \,.
\end{split}
\end{equation}
Easily to see, $U_{++}(t_2,t_1)$ is ill-defined in the neighborhood
of $-\infty$, so is $U_{--}(t_2,t_1)$ in the
neighborhood of $+\infty$. Thus, we have three classes of
evolution operators that cover the whole parameter space $[-\infty,+\infty]$, and
each class has its own singularity.
The S-operator in~\eqref{eq:92} can be constructed as
\begin{equation}
\label{eq:97}
\begin{split}
S_{+-}=U_{+-}(+\infty,-\infty)=U_{++}(+\infty,t)U_{+-}(t,t)U_{--}(t,-\infty)\,,
\end{split}
\end{equation}
where $U_{\pm\pm}$ are in their well-defined domains
respectively, but the singular $U_{+-}(t,t)$ occurs inevitably.
To avoid this singularity, another singular operator $U_{-+}(t,t)$
should be included, so that they cancel with each other
\begin{equation}
\label{eq:98}
\begin{split}
U_{+-}(t,t)U_{-+}(t,t) &= U_{-+}(t,t)U_{+-}(t,t)=I \,.
\end{split}
\end{equation}
Notice that it is the asymptotic condition
$H\stackrel{t\rightarrow\pm \infty}{\longrightarrow} H^{\pm}$ in~\eqref{eq:85} determines $U_{+-}$. Hence to obtain $U_{-+}$, another
asymptotic condition $H\stackrel{t\rightarrow\pm
\infty}{\longrightarrow} H^{\mp}$ should be added. By
combining these two conditions, we then obtain an ordinary one, either $H\stackrel{t\rightarrow\pm
\infty}{\longrightarrow} H^{+}$ or $H\stackrel{t\rightarrow\pm
\infty}{\longrightarrow} H^{-}$.

Let's then discuss the symmetries of the S-matrix of the model~\eqref{eq:82}. The only symmetries are those about the time, i.e. time translation
and time-reversal
\begin{equation}
\label{eq:99}
\begin{split}
t\rightarrow t+t_0,\qquad t\rightarrow -t \,.
\end{split}
\end{equation}
We consider only the time translation. According to
QFT~\cite{m}, a theory is invariant under time translation, it
means the same operator $e^{-iH t_0}$ acts on both the ``out''
and ``in'' states
\begin{equation}
\label{eq:100}
\begin{split}
S_{\beta\alpha} &=
\langle\Phi_{\beta}^{out}\arrowvert\Phi_{\alpha}^{in}\rangle =
\langle e^{-iH t_0}\Phi_{\beta}^{out}\arrowvert e^{-iH
t_0}\Phi_{\alpha}^{in}\rangle \,.
\end{split}
\end{equation}
It's convenient to work with the
S-operator formula. For an ordinary asymptotic
condition $H\stackrel{t\rightarrow\pm \infty}{\longrightarrow}H_0$, an operator $e^{-iH_{0}t_0}$ can be applied to act on the free state space
\begin{equation}
\label{eq:101}
\begin{split}
\langle e^{-iH_{0}t_0}\phi\arrowvert S\arrowvert
e^{-iH_{0}t_0}\chi\rangle &= \langle\phi\arrowvert e^{+iH_{0}
t_0}Se^{-iH_{0}t_0}\arrowvert\chi\rangle \,,
\end{split}
\end{equation}
where $\arrowvert\phi\rangle$ and $\arrowvert\chi\rangle$ are two
arbitrary states in the free space determined by $H_0$. Thus~\eqref{eq:100} will hold if~\cite{m}
\begin{equation}
\label{eq:102}
\begin{split}
e^{+iH_{0} t_0}Se^{-iH_{0}t_0} &= S \,.
\end{split}
\end{equation}
This can also be rewritten as
\begin{equation}
\label{eq:103}
\begin{split}
e^{+iH_{0} t_0}\Omega(+\infty)^{\dag}\Omega(-\infty)e^{-iH_{0}t_0}
&= \Omega(+\infty)^{\dag}e^{+iH t_0}e^{-iH t_0}\Omega(-\infty) \,,
\end{split}
\end{equation}
then we have~\cite{m}
\begin{equation}
\label{eq:104}
\begin{split}
e^{-iH t_0}\Omega(\pm\infty) &= \Omega(\pm\infty)e^{-iH_{0} t_0} \,,
\end{split}
\end{equation}
that is, a time translation $e^{-iH_{0}t_0}$ in the free space can induce $e^{-iH t_0}$ in the full space.
Condition~\eqref{eq:104} can be easily satisfied because of the
asymptotic condition $H\stackrel{t\rightarrow\pm
\infty}{\longrightarrow}H_0$.

However, all these don't hold for $S_{+-}$ or
$U_{+-}(+\infty,-\infty)$. In the two stationary regions
$t\rightarrow\pm \infty$, the harmonic oscillator are free, and can
be quantized as follows
\begin{equation}
\label{eq:105}
\begin{split}
\{a_{+}^\dag,a_{+},\arrowvert0^+\rangle\}(\omega_+), \qquad
\{a_{-}^\dag,a_{-},\arrowvert0^-\rangle\}(\omega_-) \,,
\end{split}
\end{equation}
with the Bogoliubov transformations~\cite{j}
\begin{equation}
\label{eq:106}
\begin{split}
a_{-}=\alpha a_{+}+\beta^* a_{+}^{\dag},\qquad a_{+}=\alpha^*
a_{-}-\beta^* a_{-}^{\dag} \,.
\end{split}
\end{equation}
These relations can lead to some unusual effects, for example, the particle
creation
\begin{equation}
\label{eq:107}
\begin{split}
N_+ = \langle0^-\arrowvert a_{+}^\dag
a_{+}\arrowvert0^-\rangle=|\beta|^2\neq 0 \,.
\end{split}
\end{equation}
The two vacua are related as~\cite{j}
\begin{equation}
\label{eq:108}
\begin{split}
\arrowvert0^-\rangle &=
\langle0^+\arrowvert0^-\rangle\exp(-\frac{1}{2}\beta^*\alpha^{-1}a_{+}^\dag
a_{+}^\dag)\arrowvert0^+\rangle \,.
\end{split}
\end{equation}
Consider the S-matrix element $\langle0^+\arrowvert
S_{+-}\arrowvert0^-\rangle$. A time translation operator
$e^{-iH^+t_0}$ will act on both the future and past Hilbert
spaces because of the relation~\eqref{eq:108}. Following the same
steps from~\eqref{eq:101} to~\eqref{eq:104}, we will obtain the
condition
\begin{equation}
\label{eq:109}
\begin{split}
e^{-iH t_0}\Omega^{\pm}(\pm\infty) &=
\Omega^{\pm}(\pm\infty)e^{-iH^+ t_0} \,.
\end{split}
\end{equation}
For $\Omega^+(+\infty)$, this condition is satisfied, since
$H\stackrel{t\rightarrow+\infty}{\longrightarrow}H^+$, while for
$\Omega^-(-\infty)$, it can not be obeyed generally,
since $H\stackrel{t\rightarrow-\infty}{\longrightarrow}H^-$ and
$[H^+,H^-]\neq0$. Hence the time translation symmetry is broken. In fact, the model~\eqref{eq:82} is only semiclassical due to a classical potential term
$\omega^2(t)q^2(t)$, with $\omega^2(t)$ treated as an
external classical field satisfying some classical equations of
motion. Besides, this potential breaks the classical symmetries of the time, the reason for the quantum symmetry breaking.

The above broken symmetry may be recovered in the following way. The condition $e^{+iH^+ t_0}S_{+-}e^{-iH^-t_0}=S_{+-}$
can be satisfied, if we enlarge the free space by combining $H^+$ with $H^-$ constrained by a relation $[H^+,H^-]=0$. This
can be achieved by replacing the parameter $t$ in $\omega^2(t)$ with
another independent one $\mu$, i.e. $\omega^2(\mu)$. There will
thus be a collection of oscillator states parameterized by $\mu$,
with the two in~\eqref{eq:105} corresponding to the limits
$\mu\rightarrow\pm \infty$. Then instead of~\eqref{eq:106}, we have
\begin{equation}
\label{eq:110}
\begin{split}
[a_{+},a_{-}] &= 0 \,.
\end{split}
\end{equation}
The whole collection of states in the free space will then becomes
\begin{equation}
\label{eq:111}
\begin{split}
\{a_{\mu}^\dag,a_{\mu},\arrowvert0\rangle\}(\omega_{\mu}) \,.
\end{split}
\end{equation}
In this case, $[a_{\mu_1},a_{\mu_2}]= 0$, and only a single vacuum $\arrowvert0\rangle$ is needed, since the Hilbert space
of the harmonic oscillator has been enlarged, with the free Hamiltonian $H_0=\int d\mu
H_{\mu}$, and an ordinary asymptotic condition
$H\stackrel{t\rightarrow\pm \infty}{\longrightarrow} H_0$.

From the above analysis we can conclude that, for a
semiclassical system such as the model~\eqref{eq:82}, a formal S-matrix or evolution
operator can be constructed, implying that the evolution of the system is unitary.
But the evolution operator has some singularities which can not be easily
avoided. Moreover, the formal S-matrix may lose some meaningful
symmetries, for example the time translation symmetry, so it
can not be used to construct well-defined physical quantities for describing the system. This conclusion can be extended
to curved space QFT, where the S-matrix may lose some important
Lorentz invariance. Certainly, a full and nonsingular description may be given by a quantum gravity theory.

\section{ A Qubit Model of The Black Hole Evaporation}
\label{sec:qubit}

In section~\ref{sec:sum}, a qubit model of the black hole evaporation is proposed, in which an EPR pairs can be utilized to correlate two distant systems without direct local couplings. In this appendix, we give some mathematical details of this model. To resemble the black hole evaporation, the EPR pairs are chosen to be at the Bell state
\begin{equation}
\label{eq:112}
\begin{split}
\arrowvert \beta_{00}\rangle_{ab}\equiv2^{-1/2}(\arrowvert
00\rangle_{ab}+\arrowvert 11\rangle_{ab}) \,.
\end{split}
\end{equation}
The initial state of the systems $A$ and $B$ are chosen to be
$\arrowvert 1\rangle_A\arrowvert 0\rangle_B$, meaning that $A$ is
excited while $B$ is at the ground state. With $A$ regarded as
the black hole and $B$ as the detector, a state
$\arrowvert 0\rangle_A\arrowvert 1\rangle_B$ is expected to occur during the evolution,
i.e. the energy of the black hole is transferred into
the detector. The unitary operation $U$ can be chosen as the one that generates the Bell states, realized by a Hadamard gate and a subsequent controlled-NOT (or CNOT) gate. The Bell states are generated as follows~\cite{n}
\begin{equation}
\label{eq:113}
\begin{split}
\arrowvert 00\rangle\stackrel{U}{\longrightarrow}\arrowvert
\beta_{00}\rangle\equiv 2^{-1/2}(\arrowvert 00\rangle+\arrowvert
11\rangle)\qquad \arrowvert
01\rangle\stackrel{U}{\longrightarrow}\arrowvert
\beta_{01}\rangle\equiv 2^{-1/2}(\arrowvert 01\rangle+\arrowvert
10\rangle) \\ \arrowvert
10\rangle\stackrel{U}{\longrightarrow}\arrowvert
\beta_{10}\rangle\equiv 2^{-1/2}(\arrowvert 00\rangle-\arrowvert
11\rangle)\qquad \arrowvert
11\rangle\stackrel{U}{\longrightarrow}\arrowvert
\beta_{11}\rangle\equiv 2^{-1/2}(\arrowvert 01\rangle-\arrowvert
10\rangle) \,.
\end{split}
\end{equation}

The initial state of the entire system $abAB$ is then
$\arrowvert\beta_{00}\rangle_{ab}\arrowvert 1\rangle_A\arrowvert
0\rangle_B$. We evolve it via a combined unitary
operation $U_{aA}U_{bB}$ following the actions in~\eqref{eq:113}.
The first action gives
\begin{equation}
\label{eq:114}
\begin{split}
\arrowvert\beta_{00}\rangle_{ab}\arrowvert 1\rangle_A\arrowvert
0\rangle_B
\stackrel{U_{aA}U_{bB}}{\longrightarrow}2^{-1/2}(\arrowvert
00\rangle_{ab}\arrowvert 1\rangle_A\arrowvert 0\rangle_B+\arrowvert
11\rangle_{ab}\arrowvert 0\rangle_A\arrowvert 1\rangle_B) \,,
\end{split}
\end{equation}
with some correlations between $A$ and $B$. Evolve once again with the same operation
\begin{equation}
\label{eq:115}
\begin{split}
\arrowvert\beta_{00}\rangle_{ab}\arrowvert 1\rangle_A\arrowvert
0\rangle_B
\stackrel{U_{aA}^2U_{bB}^2}{\longrightarrow}2^{-1/2}(\arrowvert
\beta_{00}\rangle_{ab}\arrowvert \beta_{01}\rangle_{AB}-\arrowvert
\beta_{11}\rangle_{ab}\arrowvert \beta_{10}\rangle_{AB}) \,,
\end{split}
\end{equation}
where the correlations become complicated. It can be verified
that $\arrowvert \beta_{00}\rangle_{ab}\arrowvert
\beta_{01}\rangle_{AB}$ is invariant under further operations, while $\arrowvert \beta_{11}\rangle_{ab}\arrowvert \beta_{10}\rangle_{AB}$ continues evolving as
\begin{equation}
\label{eq:116}
\begin{split}
-\arrowvert \beta_{11}\rangle_{ab}\arrowvert \beta_{10}\rangle_{AB}
\stackrel{U_{aA}U_{bB}}{\longrightarrow}\arrowvert
\beta_{01}\rangle_{ab}\arrowvert
\beta_{11}\rangle_{AB}\stackrel{U_{aA}U_{bB}}{\longrightarrow}\arrowvert
\beta_{00}\rangle_{ab}\arrowvert \beta_{11}\rangle_{AB} \,.
\end{split}
\end{equation}
By combining~\eqref{eq:115} with~\eqref{eq:116}, we then have
\begin{equation}
\label{eq:117}
\begin{split}
\arrowvert\beta_{00}\rangle_{ab}\arrowvert 1\rangle_A\arrowvert
0\rangle_B
\stackrel{U_{aA}^4U_{bB}^4}{\longrightarrow}\arrowvert\beta_{00}\rangle_{ab}\arrowvert
0\rangle_A\arrowvert 1\rangle_B \,,
\end{split}
\end{equation}
which is the required result. That is, the energy of $A$ has been
transferred into $B$, in other words, the black hole has evaporated completely\footnote{Certainly, the real black hole evaporation is much more complicated than this qubit model.}. Further evolutions are given by
\begin{equation}
\label{eq:118}
\begin{split}
\arrowvert \beta_{00}\rangle_{ab}\arrowvert \beta_{11}\rangle_{AB}
\stackrel{U_{aA}U_{bB}}{\longrightarrow}\arrowvert
\beta_{10}\rangle_{ab}\arrowvert
\beta_{11}\rangle_{AB}\stackrel{U_{aA}U_{bB}}{\longrightarrow}\arrowvert
\beta_{11}\rangle_{ab}\arrowvert \beta_{10}\rangle_{AB} \\
\stackrel{U_{aA}U_{bB}}{\longrightarrow}-\arrowvert
\beta_{01}\rangle_{ab}\arrowvert
\beta_{11}\rangle_{AB}\stackrel{U_{aA}U_{bB}}{\longrightarrow}-\arrowvert
\beta_{00}\rangle_{ab}\arrowvert \beta_{11}\rangle_{AB} \,,
\end{split}
\end{equation}
which will lead to the initial state
$\arrowvert\beta_{00}\rangle_{ab}\arrowvert 1\rangle_A\arrowvert
0\rangle_B$, by using of~\eqref{eq:115} and~\eqref{eq:116}.
Therefore, we obtain a cyclic procedure expressed as
\begin{equation}
\label{eq:119}
\begin{split}
\arrowvert\beta_{00}\rangle_{ab}\arrowvert 1\rangle_A\arrowvert
0\rangle_B
\stackrel{U_{aA}^4U_{bB}^4}{\longrightarrow}\arrowvert\beta_{00}\rangle_{ab}\arrowvert
0\rangle_A\arrowvert
1\rangle_B\stackrel{U_{aA}^4U_{bB}^4}{\longrightarrow}\arrowvert\beta_{00}\rangle_{ab}\arrowvert
1\rangle_A\arrowvert 0\rangle_B \,,
\end{split}
\end{equation}
where the second evolution can be regarded as the formation
process of a black hole. Thus we prove the BHC (i) in a not
rigorous way. In fact,~\eqref{eq:119} provides a swap operation which swaps the states
of the two input qubits of $A$ and $B$.  Moreover,~\eqref{eq:119} can also be extended to
other computational basis states, for example $\arrowvert
0\rangle_A\arrowvert 0\rangle_B$. That is to
say, analogous to~\eqref{eq:119}, we will also have
\begin{equation}
\label{eq:120}
\begin{split}
\arrowvert\beta_{00}\rangle_{ab}\arrowvert 0\rangle_A\arrowvert
0\rangle_B
\stackrel{U_{aA}^4U_{bB}^4}{\longrightarrow}\arrowvert\beta_{00}\rangle_{ab}\arrowvert
1\rangle_A\arrowvert
1\rangle_B\stackrel{U_{aA}^4U_{bB}^4}{\longrightarrow}\arrowvert\beta_{00}\rangle_{ab}\arrowvert
0\rangle_A\arrowvert 0\rangle_B \,,
\end{split}
\end{equation}
i.e. transitions between $\arrowvert 0\rangle_A\arrowvert
0\rangle_B$ and $\arrowvert 1\rangle_A\arrowvert 1\rangle_B$. Analogously,~\eqref{eq:120} gives a (double-) NOT gate for both the systems $A$ and $B$. This can be verified generally as
\begin{equation}
\label{eq:121}
\begin{split}
\arrowvert\beta_{00}\rangle_{ab}(\alpha\arrowvert
0\rangle_A+\beta\arrowvert 1\rangle_A)(\mu\arrowvert
0\rangle_B+\nu\arrowvert 1\rangle_B)
\stackrel{U_{aA}^4U_{bB}^4}{\longrightarrow}\arrowvert\beta_{00}\rangle_{ab}(\alpha\arrowvert
1\rangle_A+\beta\arrowvert 0\rangle_A)(\mu\arrowvert
1\rangle_B+\nu\arrowvert 0\rangle_B)
\,.
\end{split}
\end{equation}
Similarly, other quantum gates can also be constructed by choosing different $U_{aA}$ and $U_{bB}$.

There is a cyclic property for the evolutions in~\eqref{eq:119} and~\eqref{eq:120}, where the initial state of the EPR pairs $\arrowvert\beta_{00}\rangle_{ab}$ is recovered for the four-action and eight-action. This is related to the group structure of the unitary operation as follows. By grouping the four computational basis states as a column vector
$(\arrowvert00\rangle,\arrowvert01\rangle,\arrowvert10\rangle,\arrowvert11\rangle)^{T}$, the unitary operation given in~\eqref{eq:113} can then be rewritten
in a matrix form
\begin{equation}
\label{eq:123}
\begin{split}
U &= \frac{1}{\sqrt2}\left(\begin{array}{c c}
I_{2\times 2} & \sigma_1 \\
I_{2\times 2} & -\sigma_1
\end{array}\right) \,,
\end{split}
\end{equation}
with $I_{2\times 2}$ the unit matrix, and $\sigma_1$ the first Pauli
matrix. It is easy to verify that the unitary matrix $U$ given in
\eqref{eq:123} satisfies
\begin{equation}
\label{eq:124}
\begin{split}
U^4 = \left(\begin{array}{c c}
\sigma_1  & 0 \\
0 & \sigma_1
\end{array}\right) \qquad U^8 = \left(\begin{array}{c c}
I_{2\times 2}  & 0 \\
0 & I_{2\times 2}
\end{array}\right)\,,
\end{split}
\end{equation}
which give the cyclic property in~\eqref{eq:119} and~\eqref{eq:120}. There are also some other unitary operations with different periods, for example
\begin{subequations}\label{eq:125}
\begin{align}
\label{eq:125:1} U_1 &= \left(\begin{array}{c c}
I_{2\times 2} & 0 \\
0 & \sigma_1
\end{array}\right)U = \frac{1}{\sqrt2}\left(\begin{array}{c c}
I_{2\times 2} & \sigma_1 \\
\sigma_1 & -I_{2\times 2}
\end{array}\right) \,,
\\
\label{eq:125:2} U_2 &= \left(\begin{array}{c c}
\sigma_1 & 0 \\
0 &I_{2\times 2}
\end{array}\right)U = \frac{1}{\sqrt2}\left(\begin{array}{c c}
\sigma_1 &  I_{2\times 2}\\
I_{2\times 2} & -\sigma_1
\end{array}\right)\,,
\end{align}
\end{subequations}
both of which have a period of 2, not 8.

This qubit model also provides quantum teleportation between $A$ and $B$ by means of the EPR pairs, which can be seen as follows. Suppose we deliver a qubit $\alpha\arrowvert
0\rangle_A+\beta\arrowvert 1\rangle_A$ from $A$ to $B$, and set the initial state of $B$ to be $\arrowvert
0\rangle_B$, i.e. the initial state of the entire system is
\begin{equation}
\label{eq:122}
\begin{split}
\arrowvert\beta_{00}\rangle_{ab}(\alpha\arrowvert
0\rangle_A+\beta\arrowvert 1\rangle_A)\arrowvert 0\rangle_B \,.
\end{split}
\end{equation}
As in the ordinary quantum teleportation~\cite{n,o}, we first send $aA$ through a CNOT gate, with $A$ as the control qubit
\begin{equation}
\label{eq:122a}
\begin{split}
\stackrel{CNOT,A}{\longrightarrow}(\alpha\arrowvert\beta_{00}\rangle_{ab}\arrowvert
0\rangle_A+\beta\arrowvert\beta_{01}\rangle_{ab}\arrowvert 1\rangle_A)\arrowvert 0\rangle_B \,,
\end{split}
\end{equation}
then send $A$ through a Hadamard gate, obtaining
\begin{equation}
\label{eq:122b}
\begin{split}
\stackrel{Hadamard,A}{\longrightarrow}\{\arrowvert00\rangle_{aA}(\alpha\arrowvert
0\rangle_b+\beta\arrowvert 1\rangle_b)+\arrowvert01\rangle_{aA}(\alpha\arrowvert
0\rangle_b-\beta\arrowvert 1\rangle_b)\\+\arrowvert10\rangle_{aA}(\alpha\arrowvert
1\rangle_b+\beta\arrowvert 0\rangle_b)+\arrowvert11\rangle_{aA}(\alpha\arrowvert
1\rangle_b-\beta\arrowvert 0\rangle_b)\}\arrowvert 0\rangle_B \,,
\end{split}
\end{equation}
which is the result of the ordinary teleportation. If we make some further operations on the $bB$, the task can be accomplished. The operations are as follows
\begin{equation}
\label{eq:122c}
\begin{split}
\stackrel{Hadamard,B}{\longrightarrow}\cdots\stackrel{CNOT,B}{\longrightarrow}2^{-1}
\{\arrowvert\beta_{00}\rangle_{ab}\arrowvert 0\rangle_A(\alpha\arrowvert
0\rangle_B+\beta\arrowvert 1\rangle_B)+\arrowvert\beta_{01}\rangle_{ab}\arrowvert 0\rangle_A(\alpha\arrowvert
1\rangle_B+\beta\arrowvert 0\rangle_B)\\+\arrowvert\beta_{00}\rangle_{ab}\arrowvert 1\rangle_A(\alpha\arrowvert
0\rangle_B-\beta\arrowvert 1\rangle_B)+\arrowvert\beta_{01}\rangle_{ab}\arrowvert 1\rangle_A(\alpha\arrowvert
1\rangle_B-\beta\arrowvert 0\rangle_B)\} \,.
\end{split}
\end{equation}
Analogous to the ordinary teleportation, to complete the teleportation, some measurements on $A$ and $ab$, and some classical channel are necessary to determine the final state of $B$. This means that the information can not be transferred faster than light. In the case of the black hole, the classical channels are restricted by the classical causality, which always leads to correlation information loss in the framework of effective field theory, as discussed below~\eqref{eq:f}\footnote{When combined with the white hole in which everything interior must be emitted out of the event horizon, the above causality restriction for the black hole may disappear. In this sense, the black hole and white hole are complementary to each other. In fact, in our model developed in sections~\ref{sec:Model} and~\ref{sec:gra}, the emission and absorption parts of the interactions must be combined together to guarantee the Hermitian of the Hamiltonian, or the unitary property of the evolution. From these considerations, it seems that the black hole and white hole are actually two faces of a unique space-time structure, at least in a quantum version. This unified structure is free of the causality restriction so that classical channels are open enough to transfer classical bits, too.}. However, what we need is just transferring the black hole's information outside generally in a (effective) unitary manner, so it is not necessary to know the transferred information. Hence, the result in~\eqref{eq:122c} is enough for our black hole evaporation model.

Notice that the above qubit model is completely different
from the one in reference~\cite{p}, where the author proved a
theorem saying that the formation and evaporation of a black hole
will always lead to mixed states or remnants. The proof is based on an
argument stating that the vacuum state $\arrowvert 0_M\rangle$ is
stable during the evaporation, in the sense that the state will not be changed. In our model, however, this
condition is relaxed by emphasizing that the entanglements implicit in the state $\arrowvert 0_M\rangle$ should not be
destroyed, but the state can be changed by local operations as discussed in
section~\ref{sec:entan}\footnote{Easily to see, our effective field model or the above qubit model make an order unity modification to Hawking's leading order result, which is argued to be impossible in reference~\cite{p}. This can be seen more precisely through the calculations in~\eqref{eq:66a} and the following discussions there.}. This can also be seen from the above qubit model, where the initial $\arrowvert \beta_{00}\rangle_{ab}$ can be
changed into other three Bell states without destroying the
entanglements. Moreover, any partial measurement performed on each
qubit of the EPR pairs should be forbidden, or else the evolution~\eqref{eq:119} will be destroyed and the established correlations between $A$ and $B$ will be lost. For the quantum teleportation in~\eqref{eq:122c}, the EPR pairs are retained during the procedure, while for the ordinary teleportation in~\eqref{eq:122b}, the EPR pairs are absent and the quantum channel may be closed. In a word, as long as the entanglements of the Bell states or those (near horizon) vacua in our black hole evaporation model are not destroyed, two causally disconnected regions can be correlated all the time.

\acknowledgments

This work is supported by the NSF of China, Grant No. 11375150.

\end{document}